\documentclass[letterpaper,11pt]{article}

\usepackage{amsmath,amsthm,amssymb,amsfonts}

\usepackage[mono=false]{libertine}
\usepackage{libertinust1math}
\usepackage[T1]{fontenc}

\usepackage[margin=1in]{geometry}
\usepackage{subfiles}
\usepackage{hyperref}
\hypersetup{colorlinks=true, citecolor=blue, linkcolor=red, urlcolor=blue}
\usepackage{enumerate}
\usepackage[shortlabels]{enumitem} % Provides the ability to easily modify the amount of white space in a list
\usepackage{graphicx}
\graphicspath{{graphics/}}
\usepackage{subcaption}
\usepackage{titling}

\usepackage{color}
\usepackage{appendix}

\newtheorem{theorem}{Theorem}[section]
\newtheorem*{theorem*}{Theorem}
\newtheorem{corollary}[theorem]{Corollary}
\newtheorem{lemma}[theorem]{Lemma}
\newtheorem*{lemma*}{Lemma}
\theoremstyle{definition}

\newtheorem*{definition*}{Definition}
\theoremstyle{remark}
\newtheorem{remark}{Remark}[section]
\newcommand{\figref}[1]{Figure~\ref{#1}}
\newcommand{\secref}[1]{Section~\ref{#1}}
\newcommand{\thmref}[1]{Theorem~\ref{#1}}
\newcommand{\lemref}[1]{Lemma~\ref{#1}}

\newcommand{\corref}[1]{Corollary~\ref{#1}}

\date{}

\begin{document}

\title{Torpid Mixing of Markov Chains for the Six-vertex Model on $\mathbb{Z}^2$}
\author{
%{Jin-Yi Cai}\thanks{Department of Computer Sciences, University of Wisconsin-Madison. Supported by NSF CCF-1714275.}
%\\ \texttt{jyc@cs.wisc.edu}
%\and
{Tianyu Liu}\thanks{Department of Computer Sciences, University of Wisconsin-Madison. Supported by NSF CCF-1714275.}\\
%\thanksmark{1}\\
\texttt{tl@cs.wisc.edu}
}

%\clearpage
\maketitle

\begin{abstract}
In this paper, we study the mixing time of two widely used Markov chain algorithms for the six-vertex model, Glauber dynamics and the directed-loop algorithm, on the square lattice $\mathbb{Z}^2$.
We prove, for the first time that, on finite regions of the square lattice these Markov chains are torpidly mixing under parameter settings in the ferroelectric phase and the anti-ferroelectric phase.
\end{abstract}

\section{Introduction}\label{sec:intro}
\documentclass[paper]{subfiles}

Introduced by Linus Pauling~\cite{doi:10.1021/ja01315a102} in 1935 to describe the properties of ice, the \emph{six-vertex model} or the \emph{ice-type model} was originally studied in statistical mechanics as an abstraction of crystal lattices with hydrogen bonds.
During the following decades, it has attracted enormous interest in many disciplines of science, and become one of the most fundamental models defined on the square lattice.
In particular, the discovery of \emph{integrability} of the six-vertex models with periodic boundary conditions was considered a milestone in statistical physics~\cite{PhysRev.162.162, PhysRevLett.18.1046, PhysRevLett.19.108, PhysRevLett.19.103, PhysRevB.2.723}.

For computational expediency and modeling purposes, physicists almost entirely focused on planar lattice models.
On the square lattice $\mathbb{Z}^2$, every vertex is connected by an edge to four ``nearest neighbors''. States of the six-vertex model on $\mathbb{Z}^2$ are orientations of the edges on the lattice satisfying the \emph{ice-rule} --- every vertex has two incoming edges and two outgoing edges, i.e., they are \emph{Eulerian orientations}. The name of six-vertex model comes from the fact that there are six ways of arranging directions of the edges around a vertex (see \figref{fig:orientations}).

\begin{figure}[h!]
\centering
\begin{subfigure}[b]{0.15\linewidth}
\centering\includegraphics[width=\linewidth]{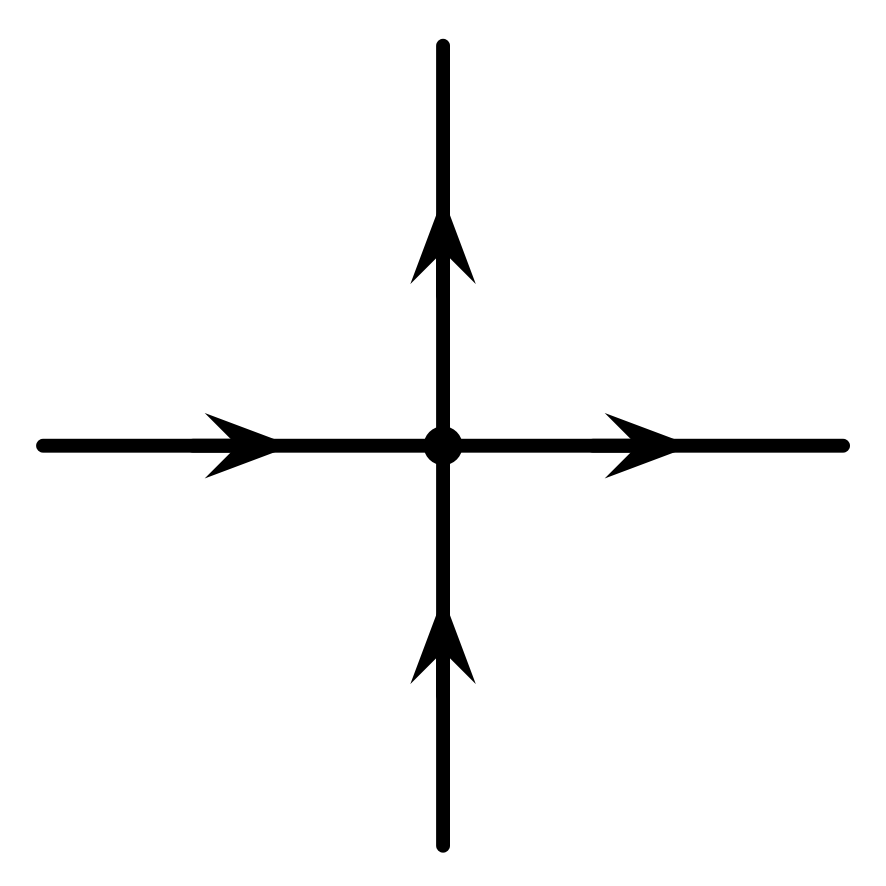}\caption{$1$}
\label{fig:orientations_1}
\end{subfigure}
\begin{subfigure}[b]{0.15\linewidth}
\centering\includegraphics[width=\linewidth]{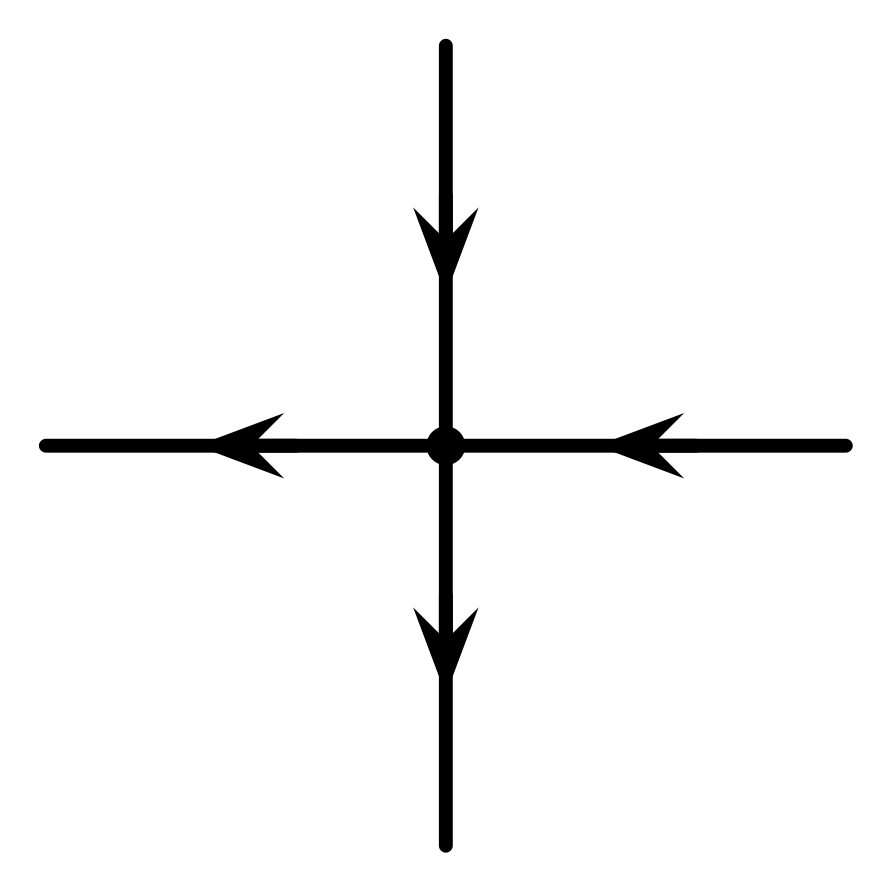}\caption{$2$}
\label{fig:orientations_2}
\end{subfigure}
\begin{subfigure}[b]{0.15\linewidth}
\centering\includegraphics[width=\linewidth]{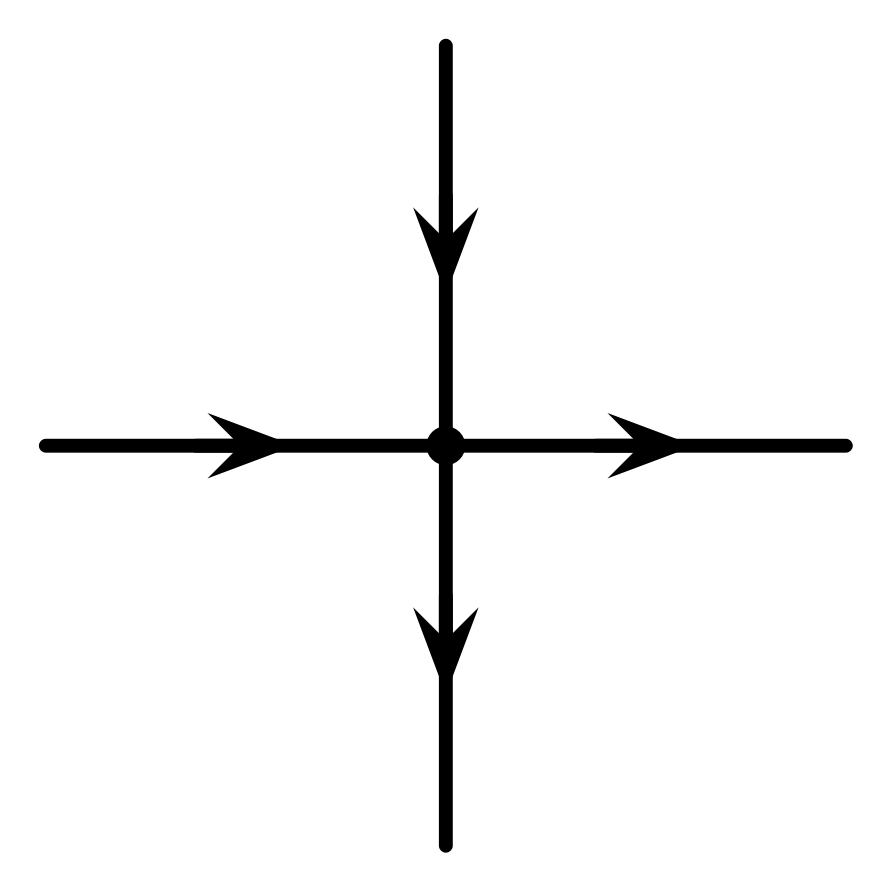}\caption{$3$}
\label{fig:orientations_3}
\end{subfigure}
\begin{subfigure}[b]{0.15\linewidth}
\centering\includegraphics[width=\linewidth]{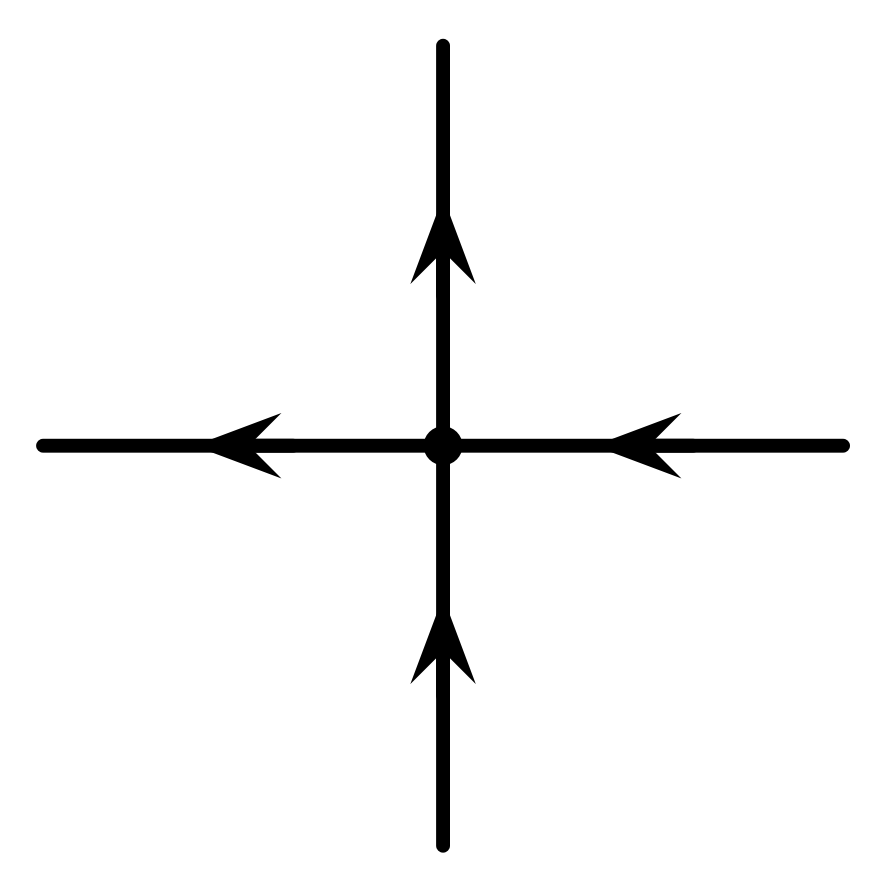}\caption{$4$}
\label{fig:orientations_4}
\end{subfigure}
\begin{subfigure}[b]{0.15\linewidth}
\centering\includegraphics[width=\linewidth]{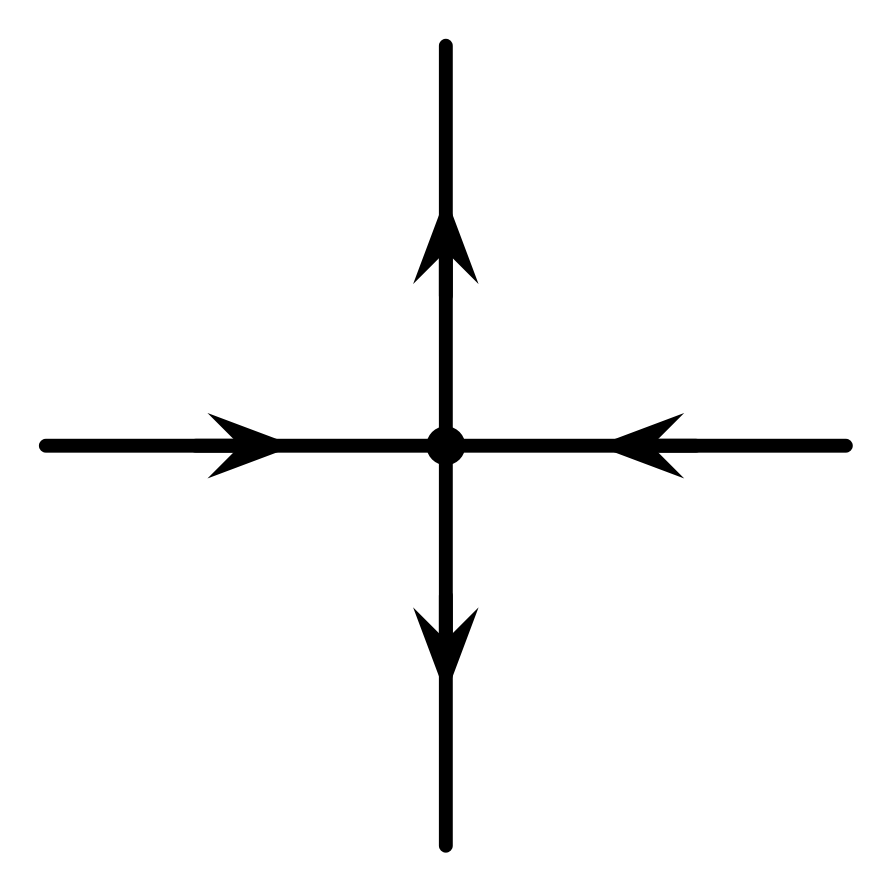}\caption{$5$}
\label{fig:orientations_5}
\end{subfigure}
\begin{subfigure}[b]{0.15\linewidth}
\centering\includegraphics[width=\linewidth]{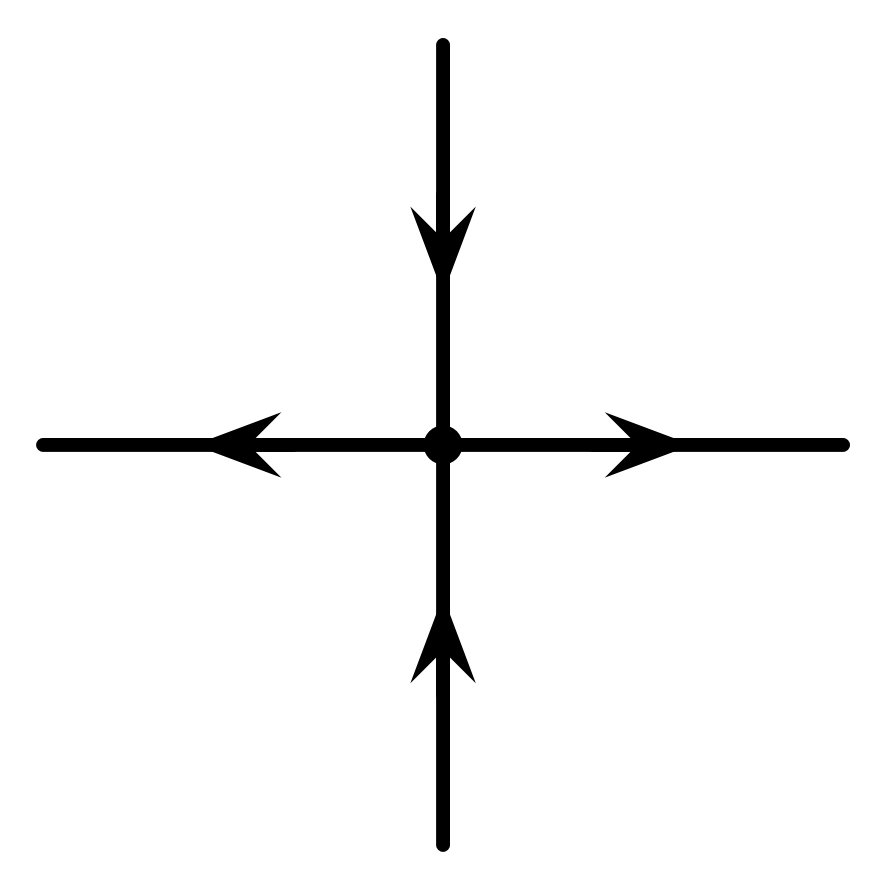}\caption{$6$}
\label{fig:orientations_6}
\end{subfigure}
\caption{Valid configurations of the six-vertex model.}\label{fig:orientations}
\end{figure}

In general, each of the six local arrangements will have a weight, denoted by $w_1, \dots, w_6$, using the ordering of \figref{fig:orientations}. The total weight of a state is the product of all vertex weights in the state.
If there is no ambient electric field, by physical considerations, then the total weight of a state should remain unchanged when flipping all arrows~\cite{Baxter:book}. Thus one may assume without loss of generality that $w_1 = w_2 = a, w_3 = w_4 = b, w_5 = w_6 = c$.
This complementary invariance is known as \emph{arrow reversal symmetry} or \emph{zero field assumption}. In this paper, we assume $a, b, c > 0$, as is the case in \emph{classical} physics.
We study the six-vertex model restricted to a finite region of the square lattice with various boundary conditions customarily studied in statistical physics literature.
On a finite subset $\Lambda \subset \mathbb{Z}^2$, denote the set of valid configurations (i.e. Eulerian orientations) by $\Omega$.
The probability that the system is in a state $\tau \in \Omega$ is given by the Gibbs distribution
\[ \mu\left(\tau\right) = \frac{1}{Z}\left(a^{n_1 + n_2} b^{n_3 + n_4} c^{n_5 + n_6}\right), \]
where $n_i$ is the number of vertices in type $i$ $(1 \le i \le 6)$ on $\Lambda$ in the state $\tau$, and the partition function $Z$ is a normalizing constant which is the sum of the weights of all states.

%The first such models were introduced by Linus Pauling \cite{doi:10.1021/ja01315a102} in 1935 to describe the properties of ice.
In 1967, Elliot Lieb~\cite{PhysRev.162.162} famously showed that,
for parameters $(a,b,c) = (1,1,1)$ on the square lattice graph,
as the side $N$ of the square approaches $\infty$,
the value of the ``partition function per vertex''
$W = Z^{1/N^2}$ approaches $\left( \frac{4}{3} \right)^{3/2}
\approx 1.5396007\ldots$ (this is called Lieb's square ice constant).
This result is called an exact solution of the model, and is considered a triumph.
After that, exact solutions for other parameter settings have been obtained in the limiting sense~\cite{PhysRevLett.18.1046, PhysRevLett.19.108, PhysRevLett.19.103, PhysRevB.2.723}.
Readers are referred to
\cite{DBLP:journals/corr/abs-1712-05880} for known results in the computational complexity of (both exactly and approximately) computing the partition function $Z$ of the six-vertex model on \emph{general} 4-regular graphs.

In statistical physics, \emph{Markov chain Monte Carlo (MCMC)} is the most popular tool to numerically study the properties of the six-vertex model. 
A partial list includes \cite{doi:10.1063/1.1678874, YANAGAWA1979329, PhysRevE.57.1155, 1999JSP....96.1091E, PhysRevE.70.016118, Allison2005, 1742-5468-2017-5-053103}.
%, PhysRevE.95.052146}.
In the literature, two Markov chain algorithms are mainly used.
The first one is \emph{Glauber dynamics}. It can be shown that there is a correspondence between Eulerian orientations of the \emph{edges} and proper three-colorings of the \emph{faces} on a rectangle region of the square lattice. (See Chapter 8 of \cite{Baxter:book} for a proof). Therefore, the Glauber dynamics for the three-coloring problem on square lattice regions (which changes a local color at each step) can be employed to sample Eulerian orientations. In fact, this simple Markov chain is used in numerical studies (e.g. in \cite{1999JSP....96.1091E, Allison2005, 1742-5468-2017-5-053103} for the density profile) of the six-vertex model under various boundary conditions.
The second one is the \emph{directed-loop algorithm}. Invented by Rahman and Stillinger~\cite{doi:10.1063/1.1678874} and widely adopted in the literature (e.g., \cite{YANAGAWA1979329, PhysRevE.57.1155, PhysRevE.70.016118}), the transitions of this algorithm are composed of creating, shifting, and merging of two ``defects'' on the edges.
An interesting aspect is that this process depicts the \emph{Bjerrum defects} happening in real ice~\cite{PhysRevE.57.1155}.
More detailed descriptions of the two Markov chain algorithms can be found in \secref{sec:prelim}.

With the heavy usage of MCMC in statistical mechanics for the six-vertex model, the efficiency of Markov chain algorithms was inevitably brought into focus by physicists.
Many of them (e.g. \cite{PhysRevE.57.1155, PhysRevE.70.016118, 1742-5468-2017-5-053103}) reported that Glauber dynamics and the directed-loop algorithms of the six-vertex model experienced significant slowdown and are even ``impractical'' for simulation purposes when the parameter settings are in the \emph{ordered phases} (see \figref{fig:diagram_phase}, in the regions FE \& AFE).
Despite the concern and numerical experience for the convergence rate of these algorithms, there is no previous provable result except for one point (that corresponds to the unweighted case) in the parameter space. This is in stark contrast to the popular studies on the mixing rate of Markov chains for the ferromagnetic Ising model~\cite{Martinelli1994I, Martinelli1994II, Cesi1996, Lubetzky2012} and hardcore gas model on lattice regions~\cite{Borgs:1999:TMM:795665.796518, Randall:2006:SMG:1109557.1109653, DBLP:conf/approx/BlancaGRT13}.

Prior to \cite{DBLP:journals/corr/abs-1712-05880}, to our best knowledge, the only provable result in the complexity of approximate sampling and counting for the six-vertex model is at the single, \emph{unweighted}, parameter setting $(a, b, c) = (1,1,1)$ where the partition function counts Eulerian orientations.
In the unweighted case, all known results are positive.
Mihail and Winkler's pioneering work \cite{Mihail1996} gave the first \textit{fully polynomial randomized approximation scheme (FPRAS)} for the number of Eulerian orientations on a general graph (not necessarily 4-regular).
Luby, Randall, and Sinclair showed that Glauber dynamics with extra moves is rapidly mixing on rectangular regions of the square lattice with fixed boundary conditions~\cite{doi:10.1137/S0097539799360355}.
Randall and Tetali proved the rapid mixing of the Glauber dynamics (without extra moves) with fixed boundary conditions by a comparison technique applied to this Markov chain and the Luby-Randall-Sinclair chain~\cite{doi:10.1063/1.533199}. Goldberg, Martin, and Paterson extended further the rapid mixing of Glauber dynamics to the free-boundary case~\cite{RSA:RSA20002}.
The unweighted setting is the single green point depicted in the blue region of \figref{fig:diagram_mixing}.

In \cite{DBLP:journals/corr/abs-1712-05880}, Cai, Liu, and Lu showed that under parameter settings $(a, b, c)$ with $a^2 \le b^2 + c^2$, $b^2 \le a^2 + c ^2$, and $c^2 \le a^2 + b^2$ (the blue region in \figref{fig:diagram_mixing}), the directed-loop algorithm mixes in polynomial time with regard to the size of input for any general 4-regular graph, resulting in an FPRAS for the partition function of the six-vertex model. Moreover, it is shown that in the ordered phases (FE \& AFE in \figref{fig:diagram_phase}), the partition function on a general graph is not efficiently approximable unless NP=RP.
Although the rapid mixing property for the directed-loop algorithm on general 4-regular graphs implies the same on the lattice region, the hardness result for general 4-regular graphs has no implications on the mixing rate of Markov chains for the six-vertex model on the square lattice in the ordered phases (FE \& AFE).

In this paper, we give the first provable negative results on mixing rates of the two Markov chains for the six-vertex model under parameter settings in the ferroelectric phases and the anti-ferroelectric phase. Our results conform to the phase transition phenomena in physics.
Here we briefly describe the phenomenon of phase transition of the 
zero-field six-vertex model (see Baxter's book~\cite{Baxter:book} for more details).
On the square lattice
in the thermodynamic limit: (1) When  $a > b + c$ (FE: ferroelectric phase)
any finite region \emph{tends to} be frozen into one of the two configurations 
where either all 
arrows point up or to the right~(Figure~\ref{fig:orientations}-1), or all point down or 
to the left~(Figure~\ref{fig:orientations}-2).
(2) Symmetrically when $b > a + c$ (also FE) all 
arrows point down or to the right~(Figure~\ref{fig:orientations}-3), or all point up or to the left~(Figure~\ref{fig:orientations}-4).
(3) When
$c > a + b$ (AFE: anti-ferroelectric phase)
configurations in Figure~\ref{fig:orientations}-5 and Figure~\ref{fig:orientations}-6 alternate.
(4) When $c < a + b$, $b < a + c$, and $a < b + c$, the system is disordered 
(DO: disordered phase)
 in the sense that all correlations decay to zero
with increasing distance; in particular
on the dashed curve $c^2 = a^2 + b^2$ the model can be solved by Pfaffians
exactly~\cite{PhysRevB.2.723}, and the correlations decay 
inverse polynomially, rather than exponentially, in distance.
See Figure~\ref{fig:diagram_phase}.

\captionsetup[subfigure]{labelformat=parens}
\begin{figure}[h!]
\centering
\begin{subfigure}[t]{0.48\linewidth}
\centering\includegraphics[width=\linewidth]{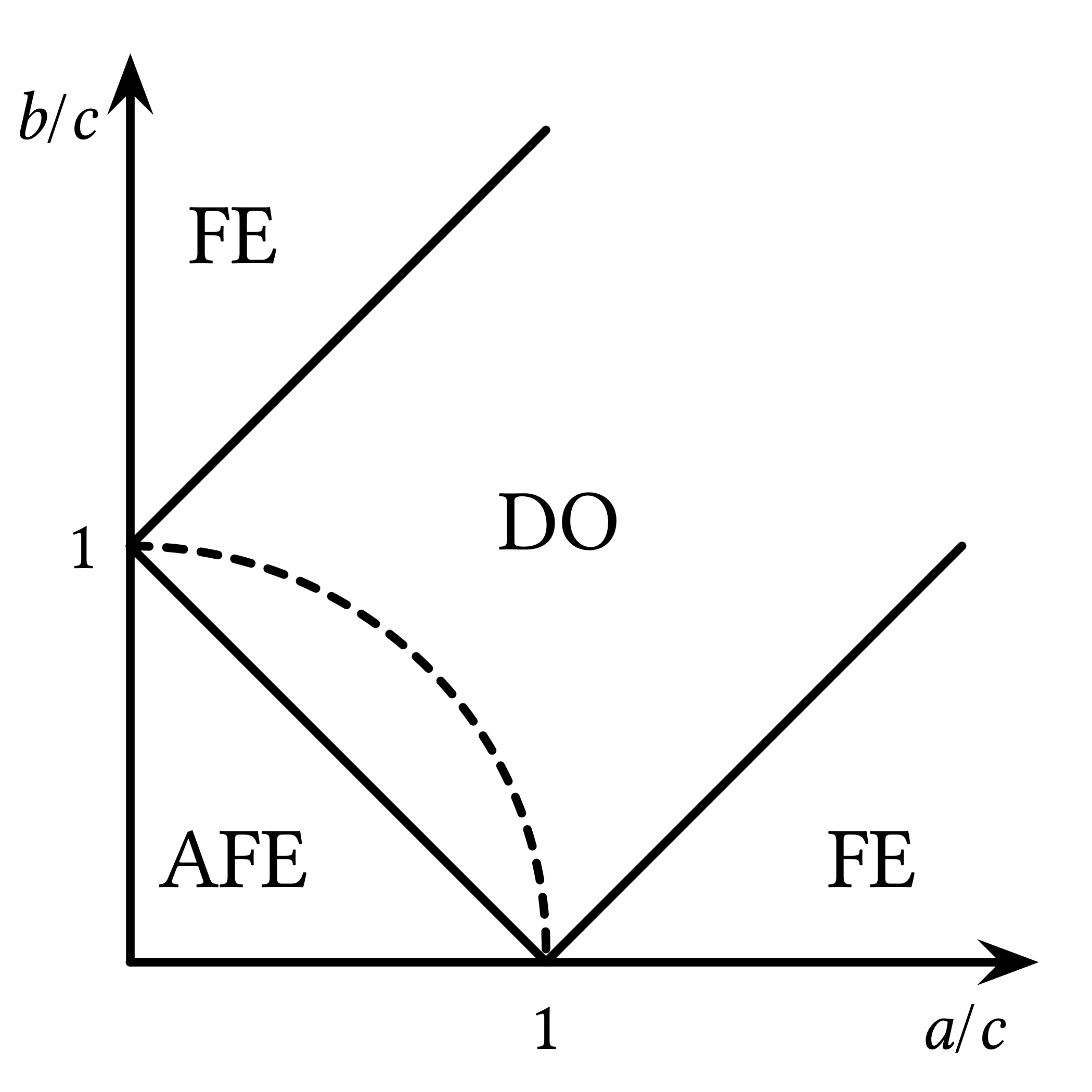}
\caption{Phase diagram of the six-vertex model.}
\label{fig:diagram_phase}
\end{subfigure}
\hfill
\begin{subfigure}[t]{0.48\linewidth}
\centering\includegraphics[width=\linewidth]{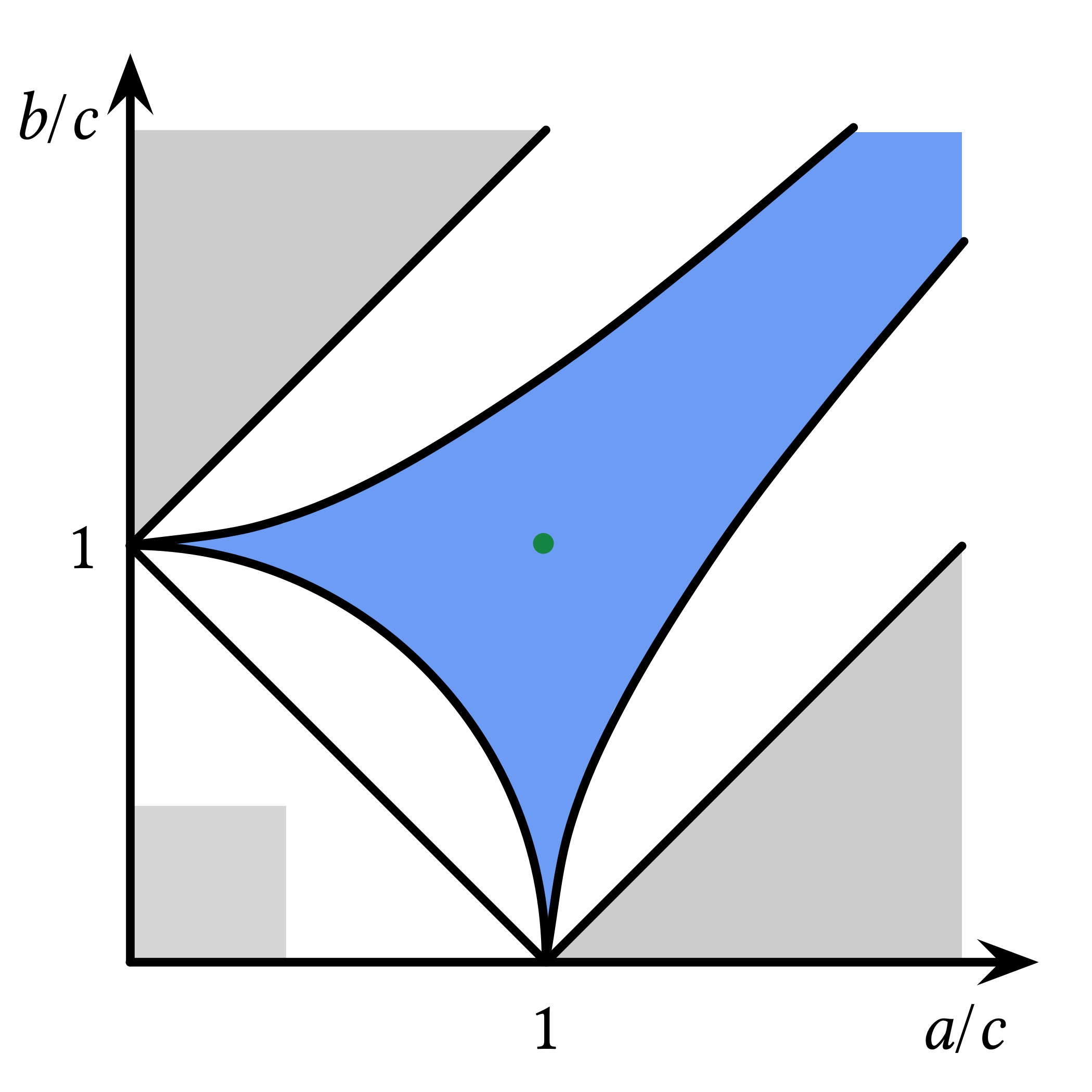}
\caption{Mixing time of Markov chains for the six-vertex model on $\mathbb{Z}^2$.}
\label{fig:diagram_mixing}
\end{subfigure}
\caption{}\label{fig:transitions}
\end{figure}

Let $\Lambda$ be a square region on the square lattice.
We show the following two theorems.
\begin{theorem}[Ferroelectric phase]\label{thm:ferro}
The directed-loop algorithm for the six-vertex model under parameter settings $(a, b, c)$ with $a > b + c$ or $b > a + c$ (i.e. the whole FE) mixes torpidly on $\Lambda$ with periodic boundary conditions.
\end{theorem}
\begin{remark}
We note that for \emph{periodic boundary conditions} Glauber dynamics is \emph{not} irreducible, so we do not consider that.
\end{remark}
\begin{theorem}[Anti-ferroelectric phase]\label{thm:anti-ferro}
Both Glauber dynamics and the directed-loop algorithm for the six-vertex model under parameter settings $(a, b, c)$ with $c \ge 2.639 \max(a,b)$ (in AFE) mix torpidly on $\Lambda$ with free boundary conditions and periodic boundary conditions.
\end{theorem}

Parameter settings covered by the above two theorems are depicted as the grey region in \figref{fig:diagram_mixing}.
Given that the \emph{$F$ model} in statistical mechanics is a special case of the six-vertex model when $a = b = 1$~\cite{PhysRevLett.18.1046}, \thmref{thm:anti-ferro} holds for the $F$ model with $c \ge 2.639$.

Our proofs build on the equivalence between small conductance and torpid mixing by Jerrum and Sinclair~\cite{SINCLAIR198993}. When arguing Markov chains for the six-vertex model in the anti-ferroelectric phase have small conductance, we switch our view between finite regions of the square lattice and their \emph{medial graphs}. This transposition allows us to adopt a \emph{Peierls argument} which has been used in statistical physics to prove the existence of phase transitions (e.g., \cite{peierls_1936, doi:10.1063/1.1666271}), and in theoretical computer science to prove the torpid mixing of Markov chains (e.g., \cite{Randall:2006:SMG:1109557.1109653, DBLP:conf/approx/BlancaGRT13}).

In the proof of \thmref{thm:anti-ferro}, we introduce a version of the \emph{fault line argument} for the six-vertex model.
Fault line arguments are introduced by Dana Randall~\cite{Randall:2006:SMG:1109557.1109653} for the lattice hardcore gas and latter adapted in \cite{LevinPeresWilmer2006} for the lattice ferromagnetic Ising, which proves torpid mixing of Markov chains via topological obstructions. The constant 2.639 comes from an upper bound for the connective constant for the square lattice self-avoiding walks~\cite{Guttmann2001}.

\section{Preliminaries}\label{sec:prelim}
\documentclass[paper]{subfiles}

\subsection{Markov chains}

\subsubsection{Glauber dynamics}
Denote by $\Lambda_n$ a square lattice region where there are $n$ vertices of degree 4 on each row and each column.
$\Lambda_n$ is in \emph{periodic boundary condition} if it forms a two-dimensional torus;
the \emph{free boundary condition} can be formulated in the following way: there are $n+2$ vertices on each row and each column, where the ``boundary vertices'' are of degree 1 and don't need to satisfy the ice-rule (and don't take weights) in a valid six-vertex configuration.
For convenience, we assume there are ``virtual edges'' connecting every two boundary vertices with unit distance on $\mathbb{Z}^2$. A virtual edge does not have orientations, serving only the purpose that every unit square inside the $(n+1) \times (n+1)$ region is closed.

Let $\Omega$ be the set of all valid configurations of the six-vertex model (Eulerian orientations) on $\Lambda_n$.
The Glauber-dynamics Markov chain, which we will denote by $\mathcal{M}_G$, has state space $\Omega$. To move from one configuration to another, this chain selects a unit square (a \emph{face}) $s$ on $\Lambda_n$ (together with the virtual edges) uniformly at random.
If all the non-virtual edges along the unit square $s$ are oriented consistently (clockwise or counter-clockwise), the chain picks a direction $d$ (clockwise or counter-clockwise) and reorients the non-virtual edges along $s$ according to the Gibbs measure.

One can easily check that such transitions take valid configurations to valid configurations.
Actually, this Markov chain is equivalent to that in \cite{RSA:RSA20002} for sampling three-colorings on the faces of $\Lambda_n$. The ergodicity of that chain translates straightforwardly to the ergodicity of $\mathcal{M}_G$ (with free boundary conditions) thanks to the equivalence between Eulerian orientations and three-colorings on $\mathbb{Z}^2$. Besides, the heat-bath move indicates that the stationary distribution of $\mathcal{M}_G$ is the Gibbs distribution for the six-vertex model.

\subsubsection{Directed-loop algorithm}
The directed-loop algorithm Markov chain, denoted by $\mathcal{M}_D$, is formally defined in \cite{DBLP:journals/corr/abs-1712-05880} for general 4-regular graphs, so here we only describe $\mathcal{M}_D$ at a high level.

The state space of $\mathcal{M}_D$ is not only $\Omega$, the ``perfect'' Eulerian orientations, but also the set of all ``near-perfect'' Eulerian orientations, denoted by $\Omega'$. For example, in \figref{fig:ferro} the state $\tau_{ur}$ is in $\Omega$ and all other five states are in $\Omega'$.
We think of each edge in $\Lambda_n$ as the two half-edges cut in the middle, and each of the half edge can be oriented independently. We say an orientation of all the half-edges is \emph{perfect} (in $\Omega$) if every pair of half-edges is oriented consistently and the ice-rule is satisfied at every vertex (except for boundary vertices under free boundary conditions); an orientation is \emph{near-perfect} (in $\Omega'$) if there are exactly two pairs of half-edges $p_1$ and $p_2$ not oriented consistently and the ice-rule is satisfied at every vertex (except for boundary vertices under free boundary conditions), with the restriction that if two half-edges in $p_1$ are oriented toward each other then in $p_2$ the two half-edges must be oriented against each other and vice versa.

The transitions in $\mathcal{M}_D$ are Metropolis moves among ``neighboring'' states. An $\Omega$ state $\tau$ and an $\Omega'$ state $\tau'$ are neighboring if $\tau'$ can be transformed from $\tau$ by picking two half-edges $e_1, e_2$ incident to a vertex $v$ with one pointing inwards $v$ and the other pointing outwards $v$ (or two half-edges $e_1, e_2$ on the boundary with one pointing towards the boundary and the other pointing against the boundary), and reverse the direction of $e_1$ and $e_2$ together. For instance, in \figref{fig:ferro} $\{\tau_{ur}, \tau_{1b}\}$ and $\{\tau_{ur}, \tau_{1c}\}$ are two pairs of neighboring states.
An $\Omega'$ state $\tau'_1$ and another $\Omega'$ state $\tau'_2$ are neighboring if $\tau'_2$ can be transformed from $\tau'_1$ by ``shifting'' one pair of conflicting half-edges one step away, while fixing the other pair of conflicting half-edges. For example, in \figref{fig:ferro} $\tau_{1c}$ and $\tau_2$ are neighboring to each other.
$\mathcal{M}_D$ can be proved to be ergodic and converges to the Gibbs measure on $\Omega \cup \Omega'$ with both free boundary conditions and periodic boundary conditions~\cite{DBLP:journals/corr/abs-1712-05880}.

\subsection{Mixing time}
The \emph{mixing time} $t_{\operatorname{mix}}$ measures the time required by a Markov chain 
to evolve to be close to its stationary distribution, in terms of \emph{total variation distance}. (The definition of mixing time can be found in \cite{LevinPeresWilmer2006}.) We say a Markov chain is torpid mixing if the mixing time is exponentially large in the input size.
A common technique to bound the mixing time is via bounding conductance, defined by Jerrum and Sinclair~\cite{SINCLAIR198993}.

Let $\pi$ denote the stationary distribution of an \emph{ergodic} and \emph{time reversible} ($\pi(x)P(x,y) = \pi(y)P(y,x)$ for any $x,y \in \Omega$) Markov chain $\mathcal{M}$ on a finite state space $\Omega$, with transition probabilities $P(x,y)$, $x, y \in \Omega$.
The \emph{conductance} of $\mathcal{M}$ is defined by
\[
\Phi = \Phi(\mathcal{M}) = \min_{\substack{S \subset \Omega\\0 < \pi(S) \le\frac{1}{2}}}\frac{Q(S,\overline{S})}{\pi(S)},
\]
where $Q(S,\overline{S})$ denotes the sum of $Q(x, y) = \pi(x)P(x,y)$ over edges in the transition graph of $\mathcal{M}$ with $x \in S$, and $y \in \overline{S} = \Omega \setminus S$.

In order to show a Markov chain mixes torpidly, we only need to prove that the conductance is (inverse) exponentially small due to the following bound~\cite{LevinPeresWilmer2006}:
\[t_{\operatorname{mix}} = t_{\operatorname{mix}}\left(\frac{1}{4}\right) \ge \frac{1}{4\Phi}.\]
As is usually assumed, Markov chains studied in this paper are all \emph{lazy} ($P(x,x) = \frac{1}{2}$ for any $x \in \Omega$) and transition probabilities ($P(x,y)$ for $x, y \in \Omega$) between neighboring states (where $P(x,y) > 0$) are at least inverse polynomially large.
Therefore, armed with the above bound, we can prove the torpid mixing of a Markov chain if we can establish the following:
\begin{enumerate}
\item
Partition the state space $\Omega$ into three subsets $\Omega_{\text{LEFT}} \cup \Omega_{\text{MIDDLE}} \cup \Omega_{\text{RIGHT}}$ as a disjoint union.
\item
Show that for any state $\tau_l \in \Omega_{\text{LEFT}}$ and $\tau_r \in \Omega_{\text{RIGHT}}$, $P(\tau_l, \tau_r) = 0$. Under the assumption that the Markov chain is irreducible (i.e., the transition graph is strongly connected), this indicates that in order to go from states in $\Omega_{\text{LEFT}}$ to states in $\Omega_{\text{RIGHT}}$, the Markov process has to go through the ``middle states'' $\Omega_{\text{MIDDLE}}$.
\item
Demonstrate that $\pi(\Omega_{\text{MIDDLE}})$ is exponentially small (compared with $\min(\pi(\Omega_{\text{LEFT}}), \pi(\Omega_{\text{RIGHT}}))$) in the input size. This means that starting from any state in $\Omega_{\text{LEFT}}$, the probability of going through $\Omega_{\text{MIDDLE}}$ (and consequently to any state in $\Omega_{\text{RIGHT}}$ and reach stationarity) is exponentially small. Hence the conclusion of torpid mixing.
\end{enumerate}

\section{Ferroelectric phase}\label{sec:ferro}
\documentclass[paper]{subfiles}

In this section we prove \thmref{thm:ferro} that $\mathcal{M}_D$ in the directed-loop algorithm for the six-vertex model in the ferroelectric phase is torpid mixing on $\Lambda_n$ with periodic boundary conditions.

For any parameter setting $(a, b, c)$ in the ferroelectric phase, either $a > b + c$ or $b > a + c$. By symmetry, without loss of generality, suppose $a > b + c$. This implies that vertex configurations as shown in \figref{fig:orientations}-1 and \figref{fig:orientations}-2 have higher weights than others. Under the periodic boundary condition, there is a state $\tau_{ur}$ in which every vertical edge points upwards and every horizontal edge points to the right (\figref{fig:ferro_ur}), i.e., every vertex on $\Lambda_n$ is in local configuration shown in \figref{fig:orientations}-1.
The total weight of $\tau_{ur}$ is $a^{n^2}$ as there are $n^2$ vertices on $\Lambda_n$.

\captionsetup[subfigure]{labelformat=parens}
\begin{figure}[h!]
\centering
\begin{subfigure}[b]{0.3\linewidth}
\centering\includegraphics[width=0.8\linewidth]{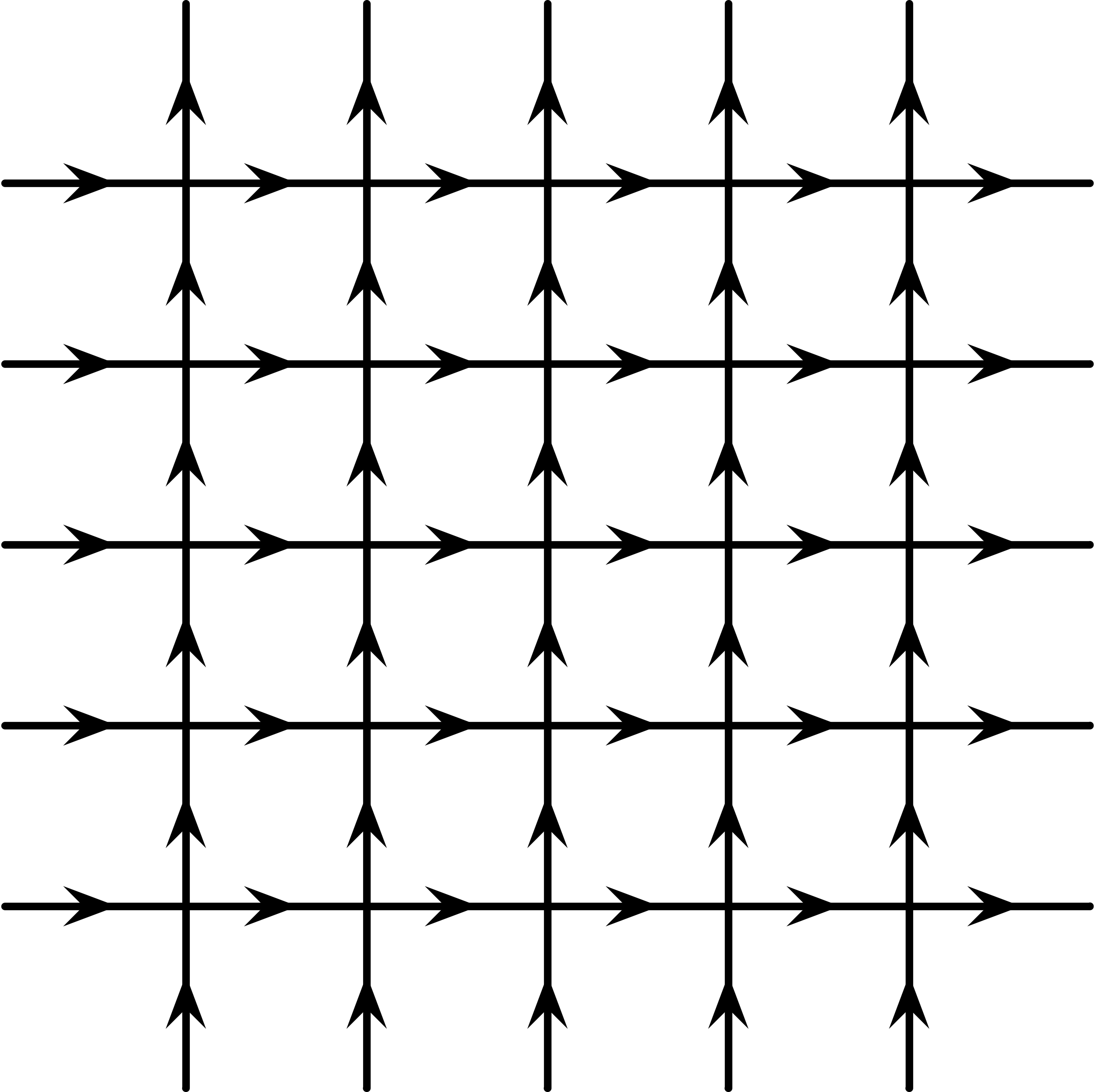}\caption{$\tau_{ur}$}\label{fig:ferro_ur}
\end{subfigure}
\begin{subfigure}[b]{0.3\linewidth}
\centering\includegraphics[width=0.8\linewidth]{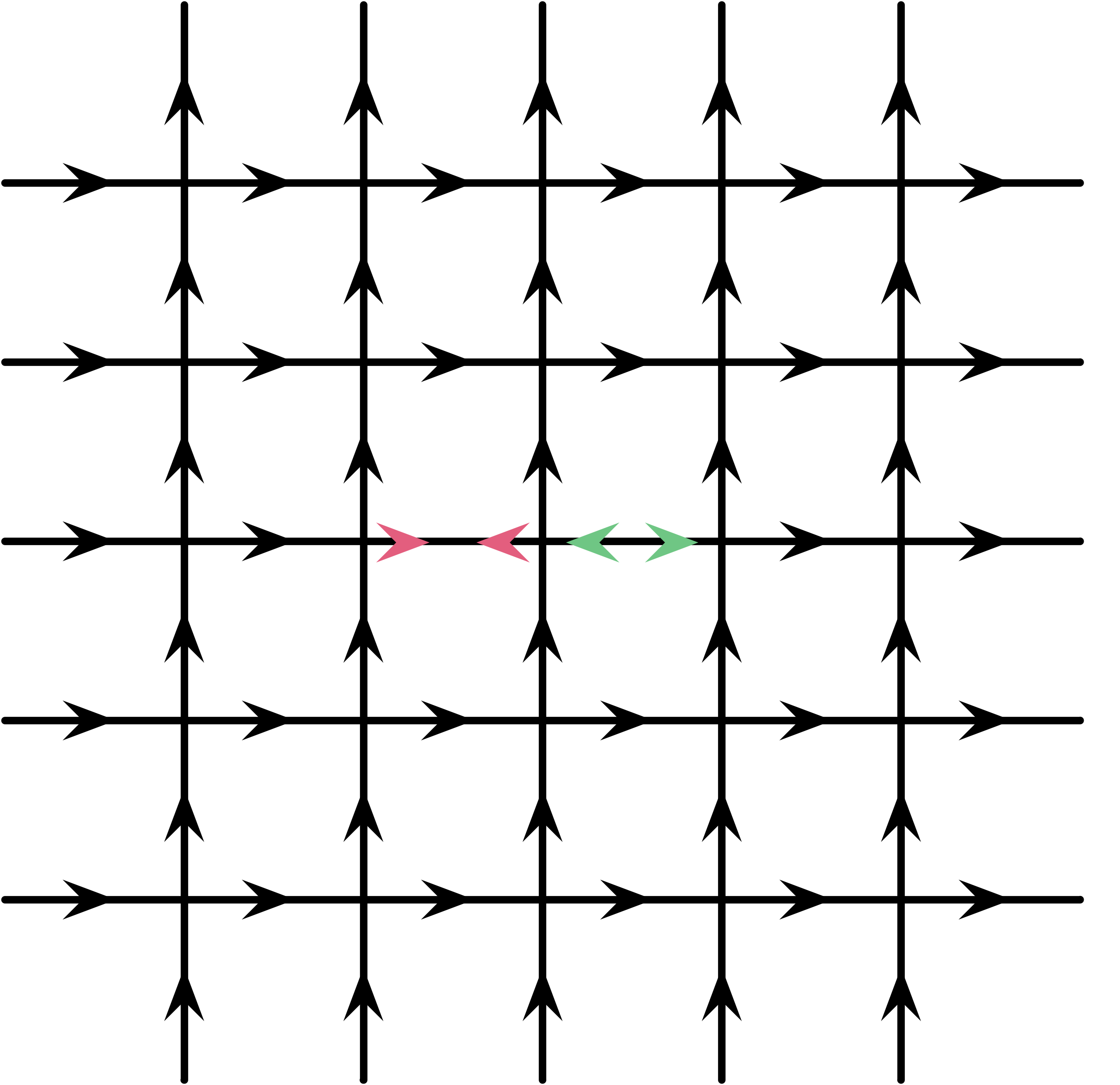}\caption{$\tau_{1b}$}
\label{fig:ferro_1b}
\end{subfigure}
\begin{subfigure}[b]{0.3\linewidth}
\centering\includegraphics[width=0.8\linewidth]{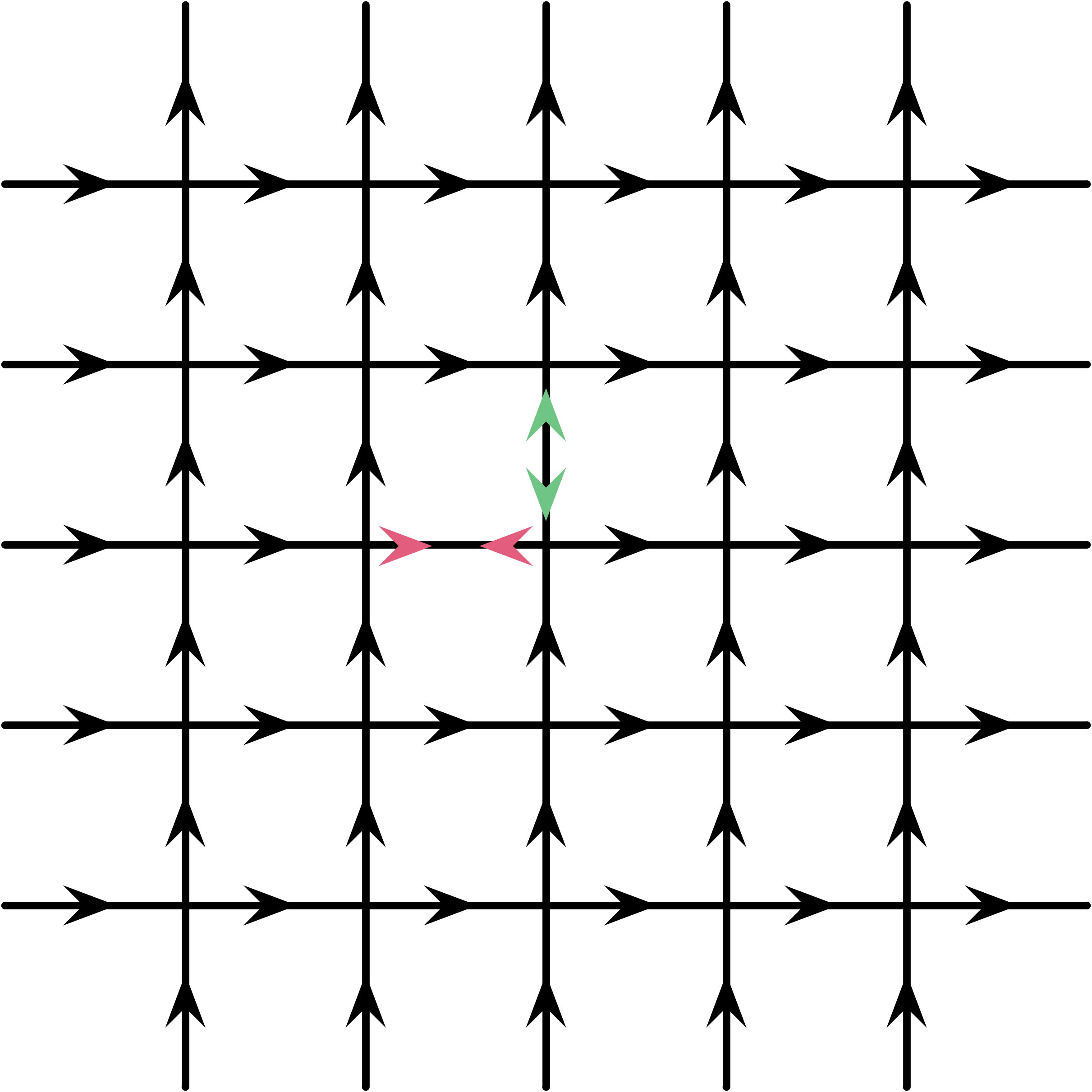}\caption{$\tau_{1c}$}
\label{fig:ferro_1c}
\end{subfigure}
\begin{subfigure}[b]{0.3\linewidth}
\centering\includegraphics[width=0.8\linewidth]{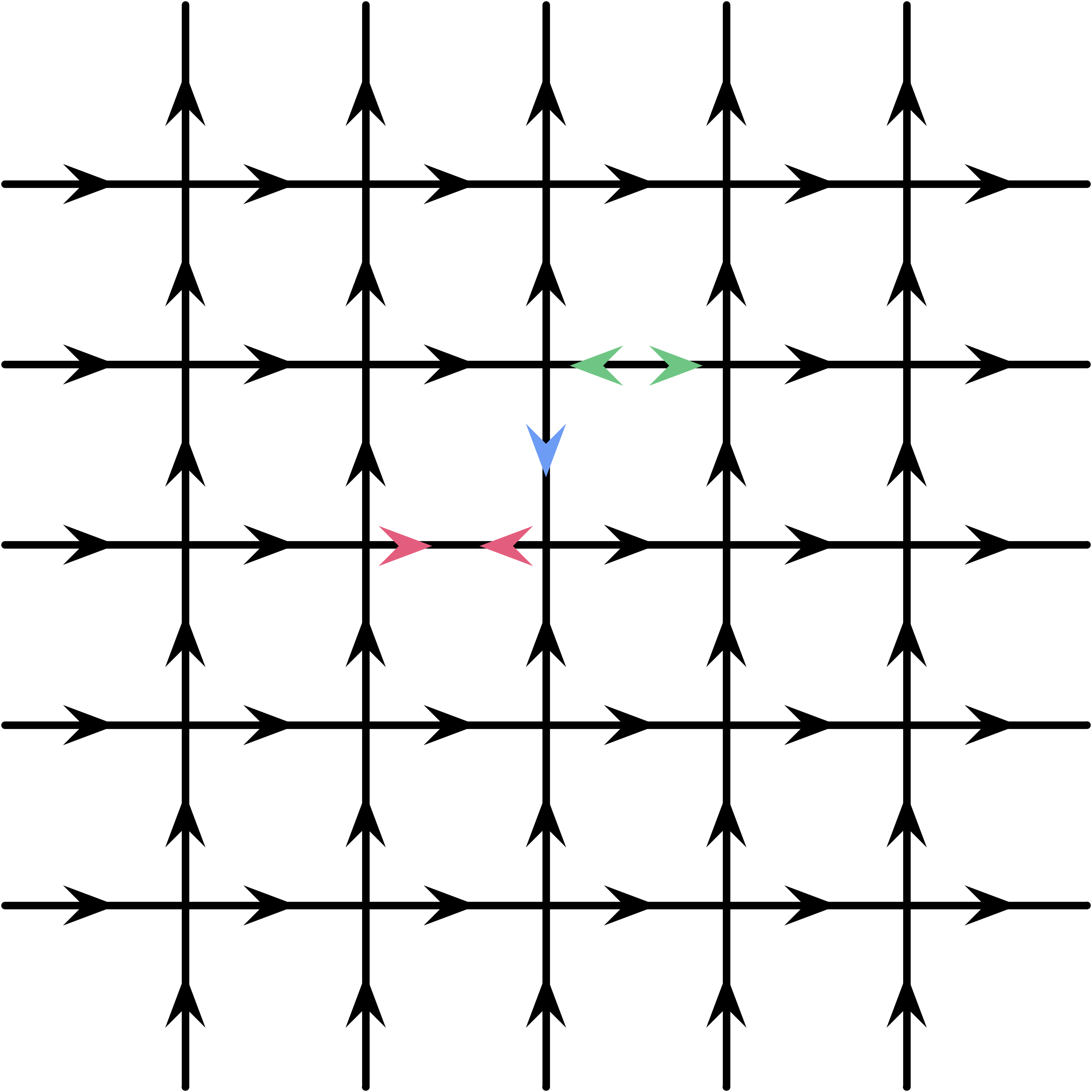}\caption{$\tau_{2}$}
\label{fig:ferro_2}
\end{subfigure}
\begin{subfigure}[b]{0.3\linewidth}
\centering\includegraphics[width=0.8\linewidth]{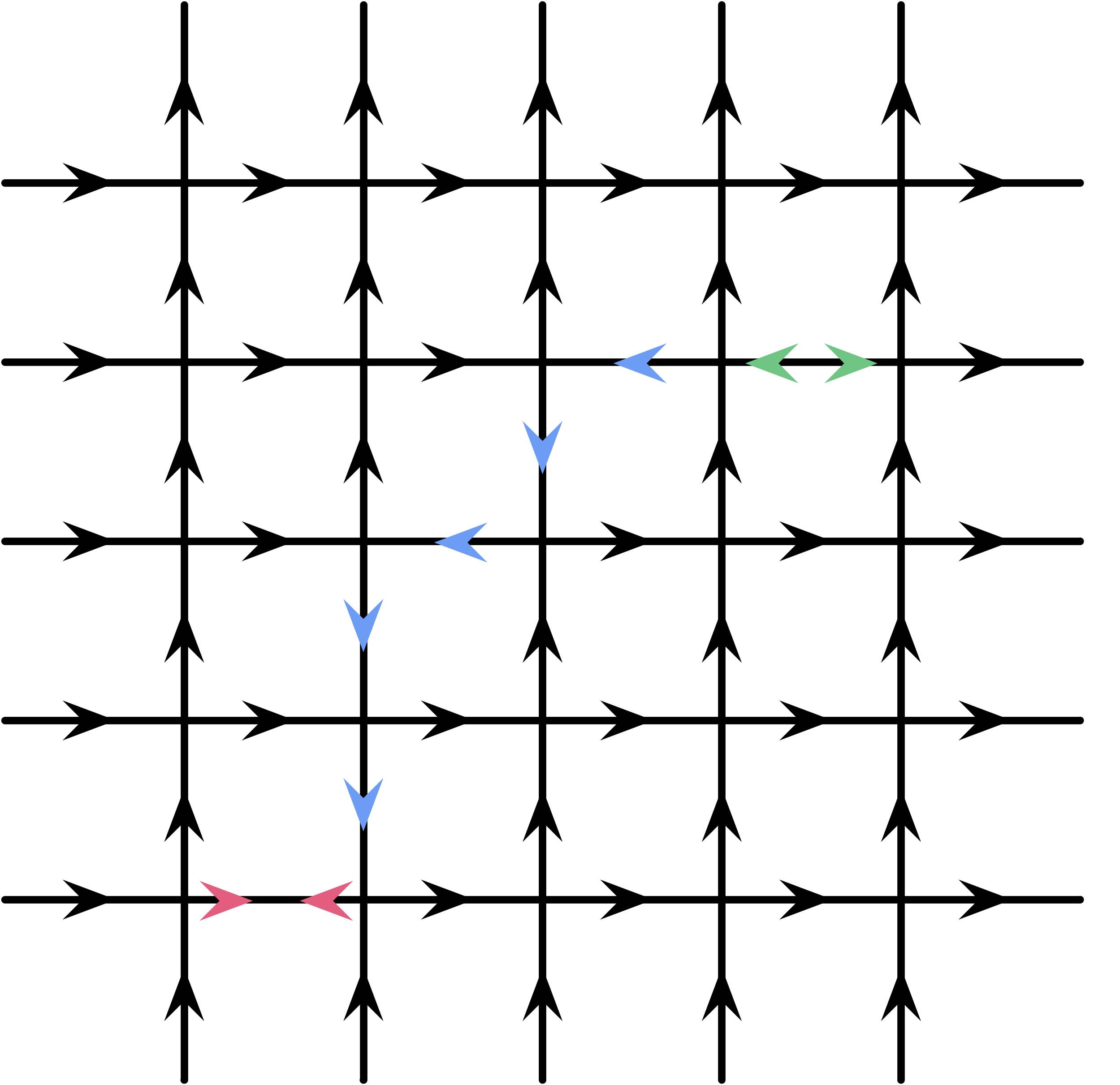}\caption{$\tau_\text{saw}$}
\label{fig:ferro_saw}
\end{subfigure}
\begin{subfigure}[b]{0.3\linewidth}
\centering\includegraphics[width=0.8\linewidth]{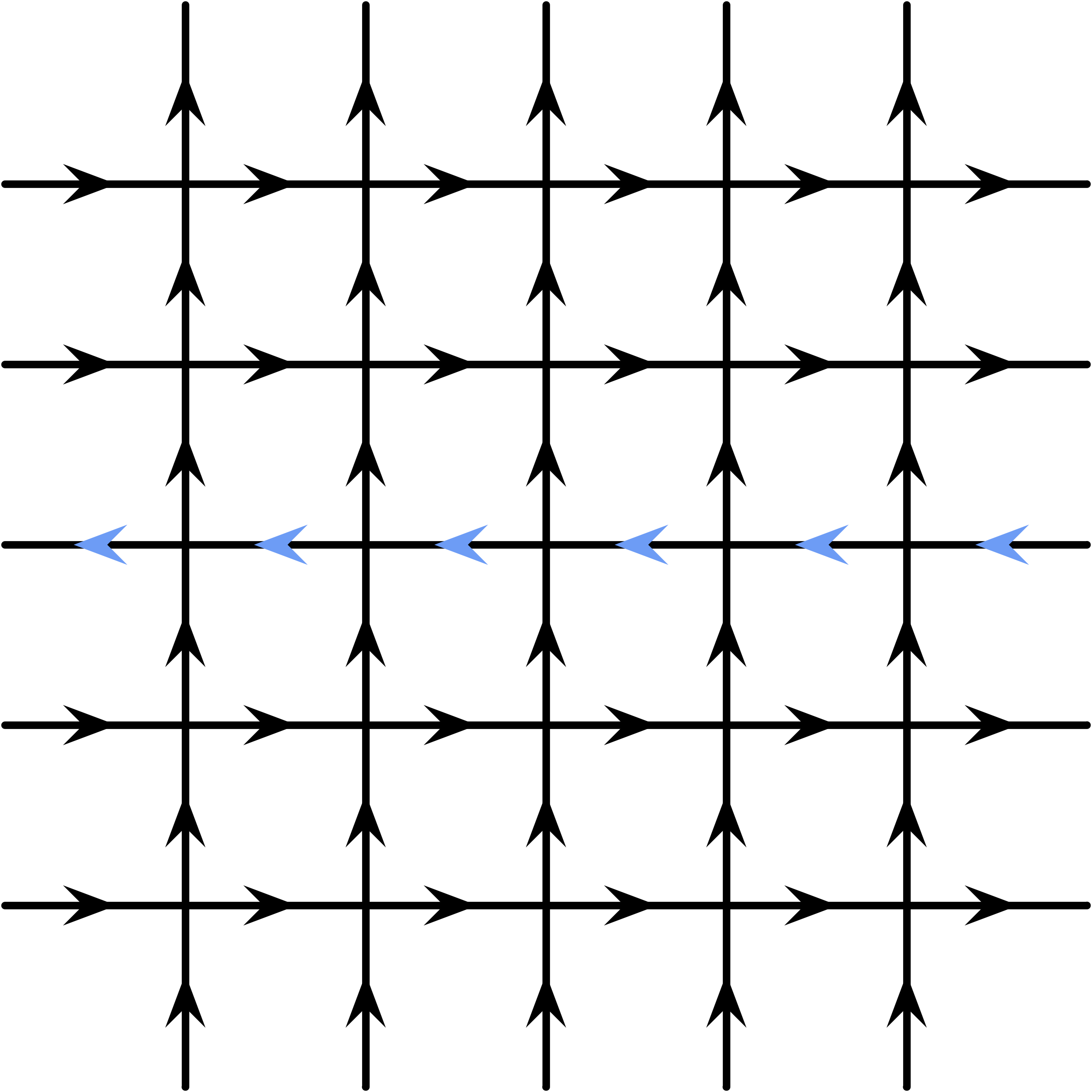}\caption{$\tau_\text{circle}$}
\label{fig:ferro_circle}
\end{subfigure}
\caption{Some states in the state space of $\mathcal{M}_D$.}\label{fig:ferro}
\end{figure}

For $\mathcal{M}_D$, the three-way partition of the state space $\Omega \cup \Omega'$ is as follows.
Denote by $T_i$ the states that can be reached from $\tau_{ur}$ in at most $i$ steps of transitions where $i$ is a nonnegative integer. Write $\partial T_i = T_i \setminus T_{i - 1}$ for $i \ge 1$. Let $\Omega_{\text{LEFT}} = T_{n-1}$, $\Omega_{\text{MIDDLE}} = \partial T_n$, and $\Omega_{\text{RIGHT}} = (\Omega \cup \Omega') \setminus (\Omega_{\text{LEFT}} \cup \Omega_{\text{MIDDLE}})$.
It is obvious that $\Omega \cup \Omega' = \Omega_{\text{LEFT}} \cup \Omega_{\text{MIDDLE}} \cup \Omega_{\text{RIGHT}}$ is a partition of the state space. Clearly $\tau_{ur} \in \Omega_{\text{LEFT}}$, thus the total weight of $\Omega_{\text{LEFT}}$ is no less than $a^{n^2}$, the weight of $\tau_{ur}$.

Before proving the total weight of $\Omega_{\text{MIDDLE}}$ is exponentially small compared with that of $\Omega_{\text{LEFT}}$ or $\Omega_{\text{RIGHT}}$, let us look at what is in $T_i$ with $0 \le i \le n$. $T_0$ is just $\{\tau_{ur}\}$.
$\partial T_1$ consists of all the states evolved from $\tau_{ur}$ by picking a vertex $v$ on $\Lambda_n$ and two incident half-edges (one pointing towards $v$ and the other away from $v$), and then reversing the orientations on these two edges. After such a transition, two pairs of conflicting half-edges are created, so $\partial T_1 \subseteq \Omega'$.

For example, the states shown in \figref{fig:ferro_1b} (state $\tau_{1b}$) and \figref{fig:ferro_1c} (state $\tau_{1c}$) are in $\partial T_1$.
The weight of $\tau_{1b}$ is $a^{n^2 - 1} b$ and that of $\tau_{1c}$ is $a^{n^2 - 1}c$.
For every state in $\partial T_1$ obtained by transitions from $\tau_{ur}$, there is exactly one vertex $v^*$ on $\Lambda_n$ no longer in the local configuration \figref{fig:orientations}-1. (Of course no vertex can be in state \figref{fig:orientations}-2.)
Actually, depending on whether the two pairs of conflicting half-edges are: (1) both vertical, (2) both horizontal, or (3) one horizontal and the other vertical, the vertex $v^*$ is in configuration shown in (1) \figref{fig:orientations}-3, (2) \figref{fig:orientations}-4, or (3) \figref{fig:orientations}-5/6, respectively.
Therefore, every state in $\partial T_1$ has weight $a^{n^2 - 1} b$ in case (1) and case (2) or $a^{n^2 - 1}c$ in case (3).

Transitions from states in $\partial T_1$ to states in $\partial T_2$ are composed of ``shifting'' one of the two conflicting pairs of half-edges to a neighboring edge on $\Lambda_n$. For example, the state in \figref{fig:ferro_2} is in $\partial T_2$.
This process will result in exactly two vertices on $\Lambda_n$ not in local configuration \figref{fig:orientations}-1 (nor in \figref{fig:orientations}-2).
As a consequence, the weight of any state in $\partial T_2$ is among $a^{n^2 - 2}b^2$, $a^{n^2 - 2}bc$, and $a^{n^2 - 2}c^2$. The state shown in \figref{fig:ferro_2} has weight $a^{n^2 - 2}c^2$.

This line of argument can be extended for $\partial T_i$ for $1 \le i \le n$, in any state of which there are exactly $i$ vertices on $\Lambda_n$ not in local configuration \figref{fig:orientations}-1 (nor in \figref{fig:orientations}-2).
When two conflicting pairs of half-edges are created in $\partial T_1$, one of them is above or to the right of another (or both). Denote the former by $p_{ur}$ (the green pair in \figref{fig:ferro}) and the latter by $p_{dl}$ (the red pair in \figref{fig:ferro}).
Observe that as the Markov chain evolves, by a single step, from a state in $\partial T_i$ to another in $\partial T_{i+1}$ (where $1 \le i \le n-1$), either $p_{ur}$ is ``pushed'' up or to the right, or $p_{dl}$ down or to the left.
For example, from \figref{fig:ferro_1c} to \figref{fig:ferro_2}, $p_{ur}$ is pushed to the right.
By induction, $p_{ur}$ is always above or to the right of $p_{dl}$ (when $i < n$).
A direct consequence is that there can be no state containing a closed circuit formed by the reversed edges (with regard to $\tau_{ur}$) in $\partial T_i$ until $i = n$. Therefore, the edges reversed in any state in $\partial T_i \ (1 \le i \le n)$ can be seen as either a \emph{self-avoiding walk} between the middle points of the two pairs of conflicting half-edges (e.g. \figref{fig:ferro_saw}) or a \emph{self-avoiding circuit} (e.g. \figref{fig:ferro_circle}). In fact, when the reversed edges form a circuit, the circuit must ``go straightforward'' at each step. This circuit is a circle parallel or perpendicular to the torus equatorial plane.
The weight of any state in $\partial T_i$ is $a^{n^2 - i} b^j c^k$ with $i = j + k$, where the values of $j$ and $k$ depend on how many ``turnarounds'' are there in the self-avoiding walk.

Therefore, the total weight of states in $\partial T_n$ is at most $n^2 \cdot a^{n^2 - n} (b + c)^n$, where $n^2$ is an upper bound on all the possible starting points for self-avoiding walks, and each monomial in $(b + c)^n$ is from a unique self-avoiding walk.
Combining with the fact that total weight of $\Omega_{\text{LEFT}}$ is at least $a^{n^2}$ (the weight of $\tau_{ur}$) and is at most that of $\Omega_{\text{RIGHT}}$ (because there is a weight-preserving injective map from $\Omega_\text{LEFT}$ to $\Omega_\text{RIGHT}$ by reversing orientations of all the edges), we know that the conductance of $\mathcal{M}_D$ is at most $\frac{n^2 \cdot a^{n^2 - n} (b + c)^n}{a^{n^2}} = n^2\left(\frac{b + c}{a}\right)^n$. This is exponentially small in $n$ since $a > b + c$ are fixed constants in the ferroelectric phase.

\section{Anti-ferroelectric phase}\label{sec:anti-ferro}
\documentclass[paper]{subfiles}

In this section, we prove the following theorem which is part of \thmref{thm:anti-ferro}.
After proving \thmref{thm:anti-ferro_sub}, we state the ideas needed to extend it to \thmref{thm:anti-ferro}, the full proof of which is omitted due to space limit.

As we did in the ferroelectric phase, the intuition behind our proof for the anti-ferroelectric phase is to find a partition $\Omega = \Omega_{\text{LEFT}} \cup \Omega_{\text{MIDDLE}} \cup \Omega_{\text{RIGHT}}$ of the state space of $\mathcal{M}_G$, i.e., all the Eulerian orientations on $\Lambda_n$.
However, the strategy is different from that used in \secref{sec:ferro} --- here the subset $\Omega_{\text{MIDDLE}}$ is determined in terms of a topological obstruction.

\begin{theorem}[Anti-ferroelectric phase]\label{thm:anti-ferro_sub}
Glauber dynamics for the six-vertex model under parameter settings (a, b, c) with $c \ge 2.639 \max(a,b)$ mix torpidly on $\Lambda_n$ with free boundary conditions.
\end{theorem}

Observe that there are two states in $\Omega$ with maximum weights: $\tau_\text{G}$ (\figref{fig:anti-ferro_green}) and $\tau_\text{R}$ (\figref{fig:anti-ferro_red}) where every vertex is in local configuration \figref{fig:orientations}-5 or \figref{fig:orientations}-6, and thus has vertex weight $c$.
Since $\tau_\text{G}$ and $\tau_\text{R}$ are total reversals of each other in edge orientations, for any edge in any state $\tau \in \Omega$, it is oriented either as in $\tau_\text{G}$ or as in $\tau_\text{R}$.
Let us call an edge to be \emph{green} if it is oriented as is in $\tau_\text{G}$ and \emph{red} otherwise.
Observe that in order to satisfy the ice-rule (2-in-2-out), the number of green (and thus also two red) edges incident to any vertex (except for the boundary vertices) is always even $(0, 2, \text{ or } 4)$, and if there are two green (or red) edges they must be \emph{rotationally adjacent} to each other. See \figref{fig:anti-ferro_eg} for an example. Also note that the four edges along a unit square on $\mathbb{Z}^2$ are all red edges or all green edges if and only if they are oriented consistently, hence flippable by a single move of $\mathcal{M}_G$.

\captionsetup[subfigure]{labelformat=parens}
\begin{figure}[h!]
\centering
\begin{subfigure}[b]{0.3\linewidth}
\centering\includegraphics[width=0.8\linewidth]{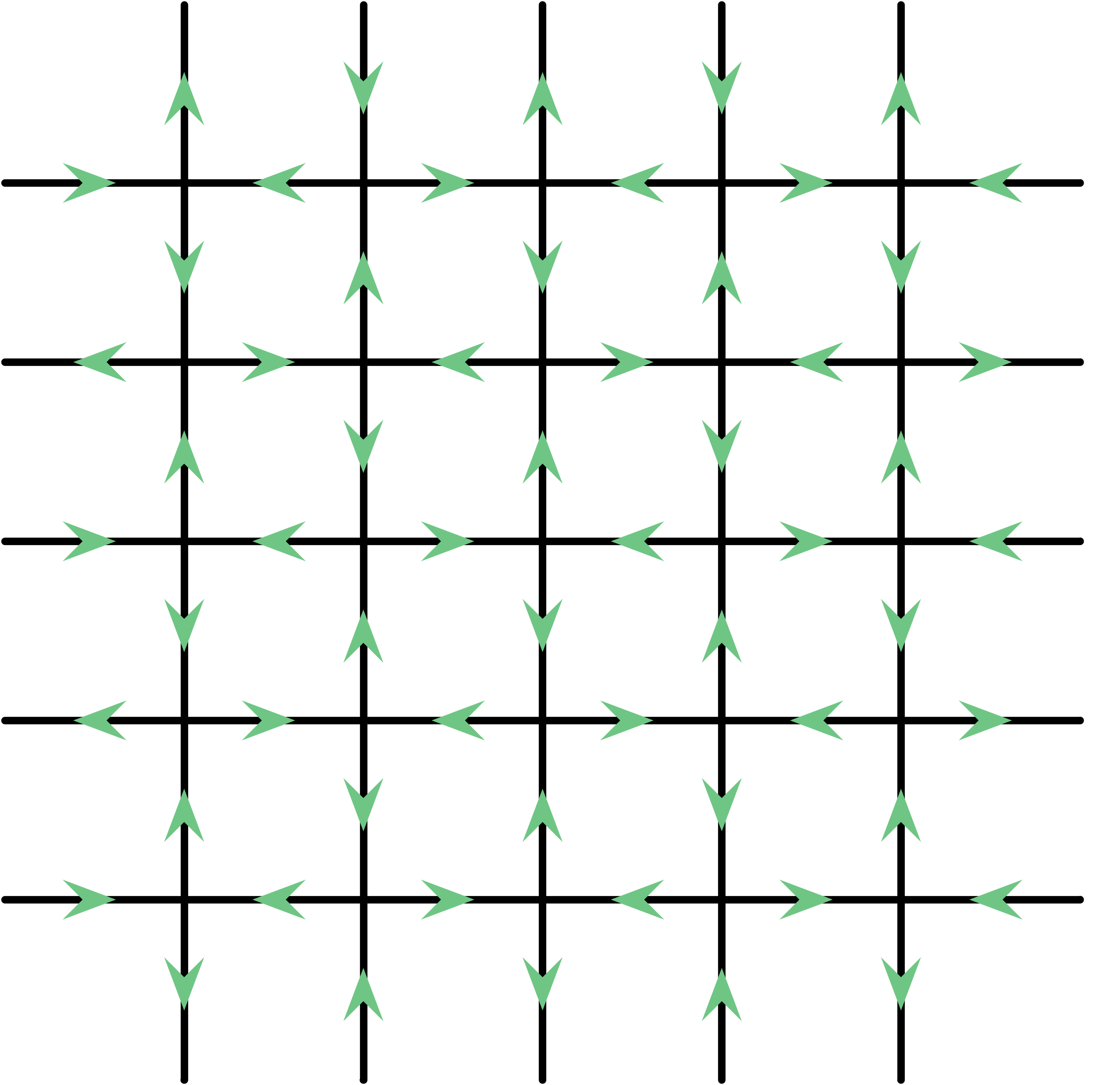}\caption{$\tau_\text{G}$}\label{fig:anti-ferro_green}
\end{subfigure}
\begin{subfigure}[b]{0.3\linewidth}
\centering\includegraphics[width=0.8\linewidth]{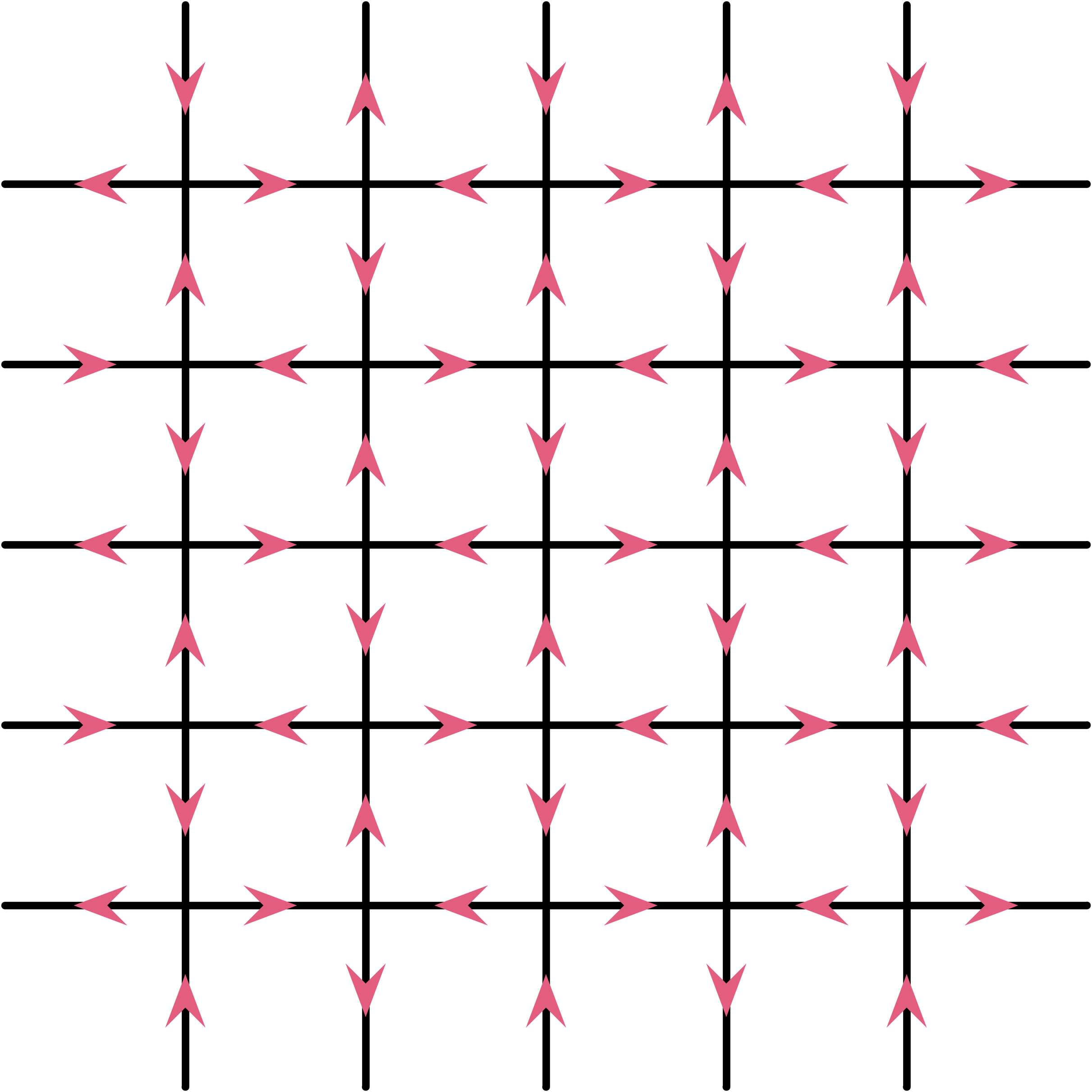}\caption{$\tau_\text{R}$}
\label{fig:anti-ferro_red}
\end{subfigure}
\begin{subfigure}[b]{0.3\linewidth}
\centering\includegraphics[width=0.8\linewidth]{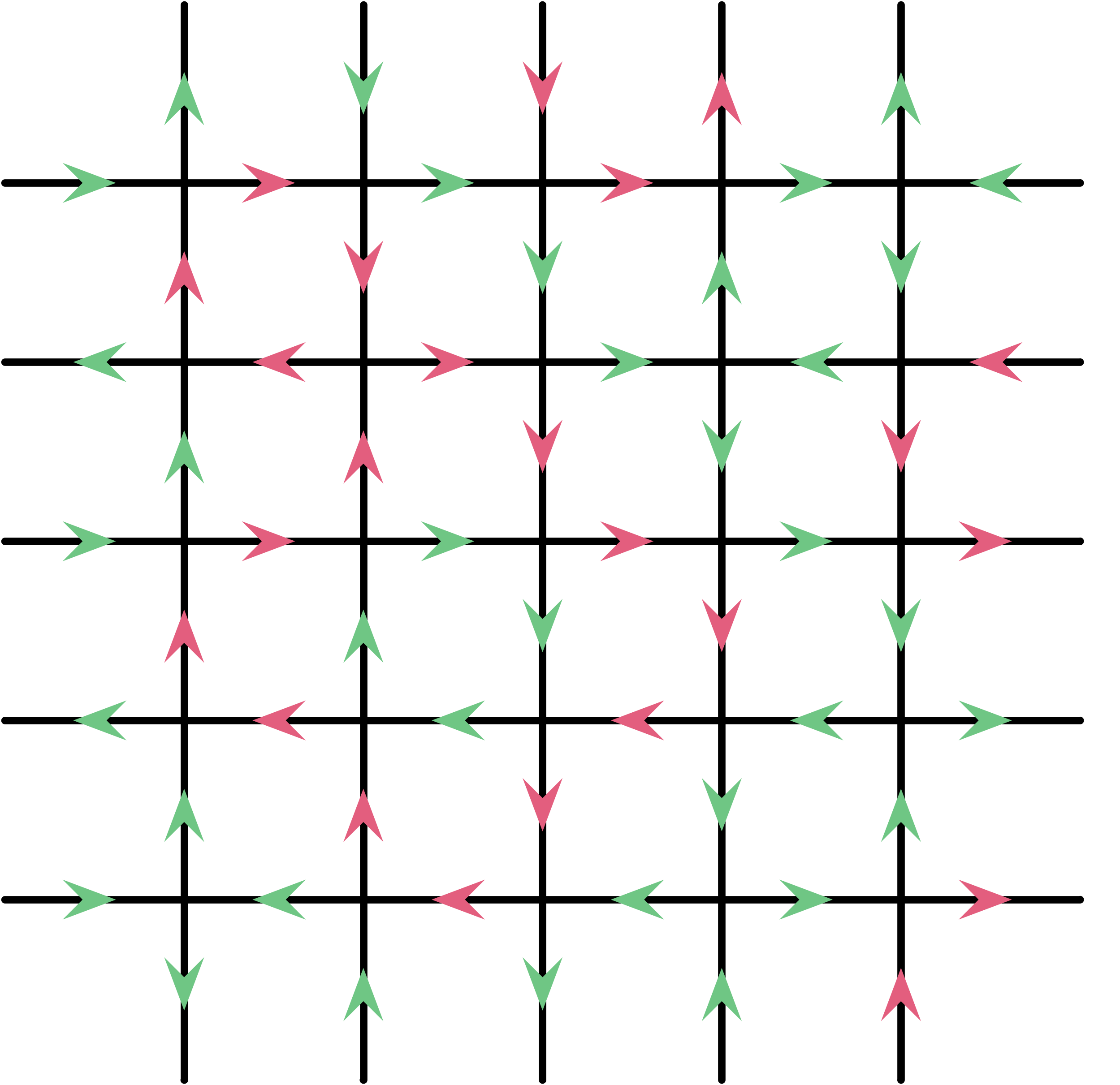}\caption{$\tau$}
\label{fig:anti-ferro_eg}
\end{subfigure}
\caption{Some states in the state space of $\mathcal{M}_G$.}\label{fig:anti-ferro}
\end{figure}

We say a simple path from a horizontal edge on the left boundary of $\Lambda_n$ to a horizontal edge on the right boundary of $\Lambda_n$ is a \emph{horizontal green (or red) bridge} if the path consists of only green (or red, respectively) edges; a \emph{vertical green (or red) bridge} is defined similarly.
A state $\tau \in \Omega$ has a \emph{green cross} if it has both a green horizontal bridge and a green vertical bridge; a \emph{red cross} is defined similarly.
Let $C_{\text{G}} \subset \Omega$ denote the states having a green cross and $C_{\text{R}}$ the states having a red cross.
In the following lemma, we prove that $C_{\text{G}} \cap C_{\text{R}} = \emptyset$.

\begin{lemma}\label{lem:anti-ferro_no_cross}
A green cross and a red cross cannot coexist.
\end{lemma}
\begin{proof}
It suffices to show that a green horizontal bridge precludes a red vertical bridge.
Consider a virtual point $v_\text{L}$ sitting to the left of $\Lambda_n$ connected by an edge to every (external) vertices of $\Lambda_n$ on the left boundary, and another virtual point $v_\text{R}$ connected by an edge to every vertex of $\Lambda_n$ on the right boundary.
Connect $v_\text{L}$ and $v_\text{R}$ by an edge below $\Lambda_n$.

If there is a green horizontal bridge, then by definition there is a continuous closed curve $\mathcal{C}$ formed by the bridge and some edges we added (\figref{fig:anti-ferro_curve}).
According to the \emph{Jordan Curve Theorem}, $\mathcal{C}$ separates the plane into two disjoint regions, the inside and the outside. Vertices of $\Lambda_n$ that are on the bottom boundary are inside; vertices on the top boundary are outside. Therefore, in order to have a red vertical bridge, there must be a simple red path going across $\mathcal{C}$. That is to say, a red vertical bridge must cross the green horizontal bridge.

\captionsetup[subfigure]{labelformat=parens}
\begin{figure}[h!]
\centering
\begin{subfigure}[b]{0.4\linewidth}
\centering\includegraphics[width=\linewidth]{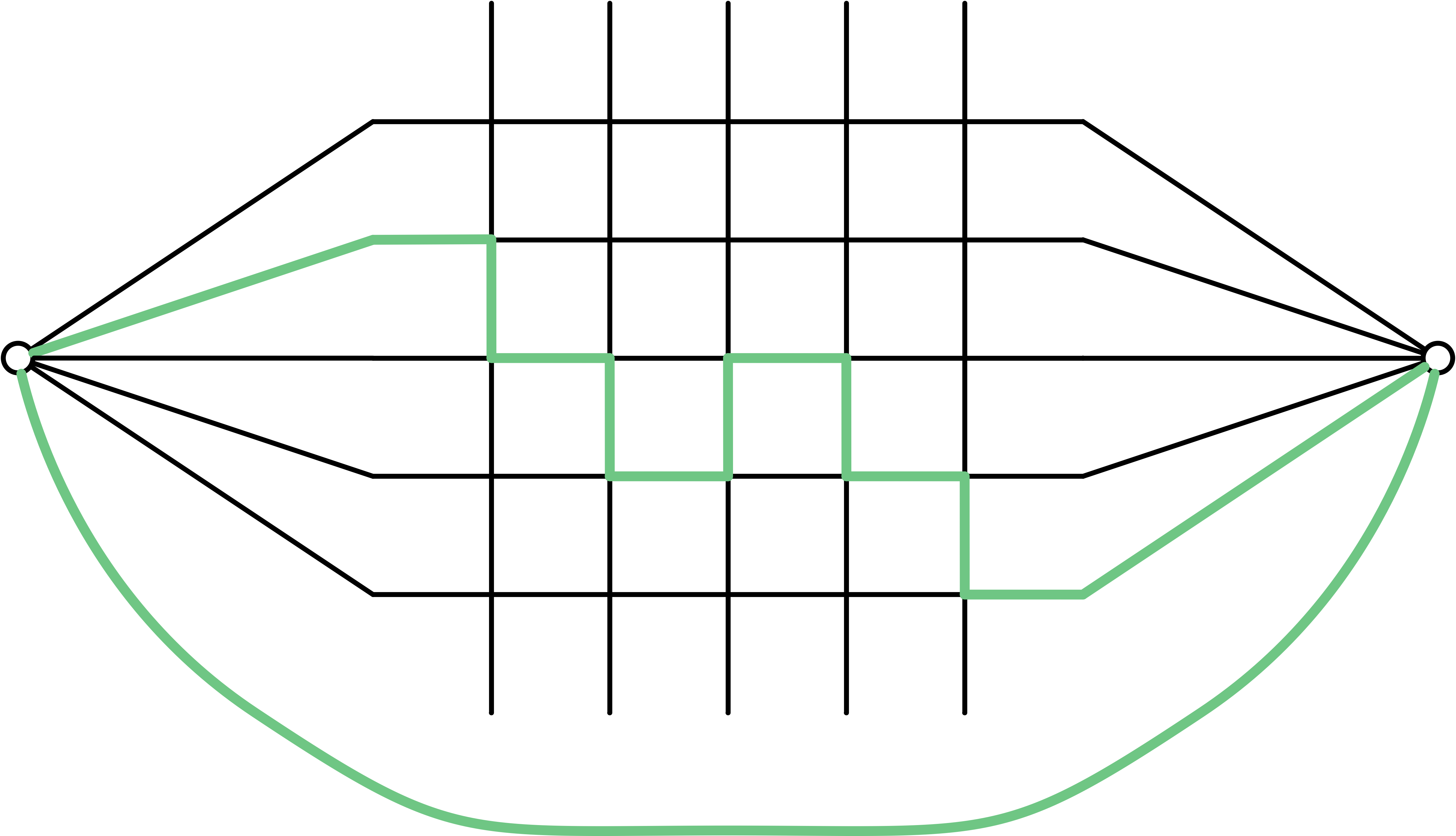}\caption{A closed curve $\mathcal{C}$.}\label{fig:anti-ferro_curve}
\end{subfigure}
\begin{subfigure}[b]{0.4\linewidth}
\centering\includegraphics[width=0.5\linewidth]{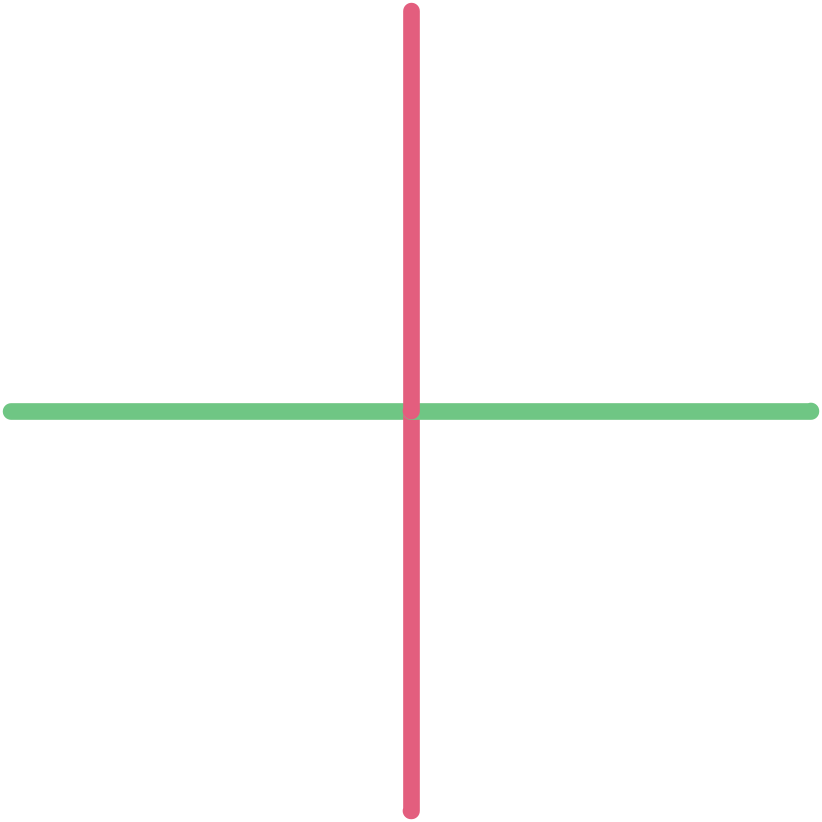}\caption{An impossible configuration.}
\label{fig:anti-ferro_cross}
\end{subfigure}
\caption{}\label{fig:anti-ferro_no_cross}
\end{figure}

However, this is impossible.
Clearly, being of different colors, a red bridge and a green bridge cannot share any edge.
Since the local configuration shown in \figref{fig:anti-ferro_cross} means that the four edges incident to a vertex $(i, j)$ on $\Lambda_n$ are all pointing inwards (when $i + j$ is even) or all pointing outwards (when $i + j$ is odd), it is not allowed in any valid states of six-vertex configurations. Similarly, the local configuration of a vertex surrounded by two red horizontal edges and two green vertical edges (a 90 degree rotation of \figref{fig:anti-ferro_cross}) is also not allowed.
\end{proof}

Next we characterize the states in $\Omega \setminus (C_{\text{G}} \cup C_{\text{R}})$.
Define a shifted lattice\footnote{Strictly speaking, a lattice is a discrete subgroup of $\mathbb{R}^n$. A shifted copy of a lattice does not contain 0.} $L$ to be $\mathbb{Z}^2 + \left(\frac{1}{2}, \frac{1}{2}\right)$ where two points $(a, b)$ and $(c, d)$ in $L$ are neighbors if $|a-c| = |b-d| = 1$ ($a$, $b$, $c$, and $d$ are all half integers), i.e., they are at the center of a square in $\mathbb{Z}^2$ and are connected by ``diagonal'' edges of length $\sqrt{2}$. An example of $L$ and its relationship with $\mathbb{Z}^2$ is shown in \figref{fig:anti-ferro_lattice}. $L$ is not connected --- it is composed of two sub-lattices $L_0$ and $L_1$ (depicted with different colors in \figref{fig:anti-ferro_lattice_sub}).
Denote by $L_n$ the restriction of $L$ on the finite region inside $\Lambda_n$.
Note that in graph theoretical terms, the square lattice $\Lambda_n$ is planar and 4-regular, and thus can be seen as the \emph{medial graph} of two planar graphs. In fact, they are $L_0$ and $L_1$ (restricted onto $L_n$).

From now on, we use $\Lambda_n$-vertices/edges as an abbreviation for vertices/edges in $\Lambda_n$; and we use $L_n$-vertices/edges and other similar notations whenever it has a clear meaning in the context.
For any state $\tau \in \Omega$, there is a subset $L_\tau$ of $L_n$-edges associated with $\tau$. Each $L_n$-edge $e$ ``goes diagonally through'' exactly one $\Lambda_n$-vertex, denoted by $v_e$. We say $e$ is in $L_\tau$ if and only if the four $\Lambda_n$-edges incident to $v_e$ are 2-green-2-red and $e$ \emph{separates} the two green edges from the two red edges (Remember that in this case edges in the same color must be rotationally adjacent to each other). See \figref{fig:anti-ferro_eg_lattice} for an instance of a state $\tau$ and its associated $L_\tau$.
In the following we abuse the notation and use $L_\tau$ as its induced subgraph of $L$.
This view was adopted by \cite{doi:10.1063/1.1666271} for establishing the existence of the spontaneous staggered polarization in the anti-ferroelectric phase of the six-vertex model.

For any $\tau \in \Omega$ and $L_\tau$, we make the following observations:
\begin{itemize}
\item
There is always an even number of $L_\tau$-edges meeting at any $L_n$-vertex, except for the $L_n$-vertices on the boundary. Because this number is equal to the number of times for the color change on the four $\Lambda_n$-edges surrounding the $L_n$-vertex, if we start from any one of the four $\Lambda_n$-edges and go rotationally over the four $\Lambda_n$-edges, which is even.
\item
For any $\Lambda_n$ vertex, there can be at most one $L_\tau$-edge going through which is either in $L_0$ or in $L_1$.
\item
If $\bar{\tau} \in \Omega$ is the state by a total edge reversal of $\tau$, then $L_{\bar{\tau}} = L_\tau$.
\end{itemize}

\captionsetup[subfigure]{labelformat=parens}
\begin{figure}[h!]
\centering
\begin{subfigure}[b]{0.3\linewidth}
\centering\includegraphics[width=0.8\linewidth]{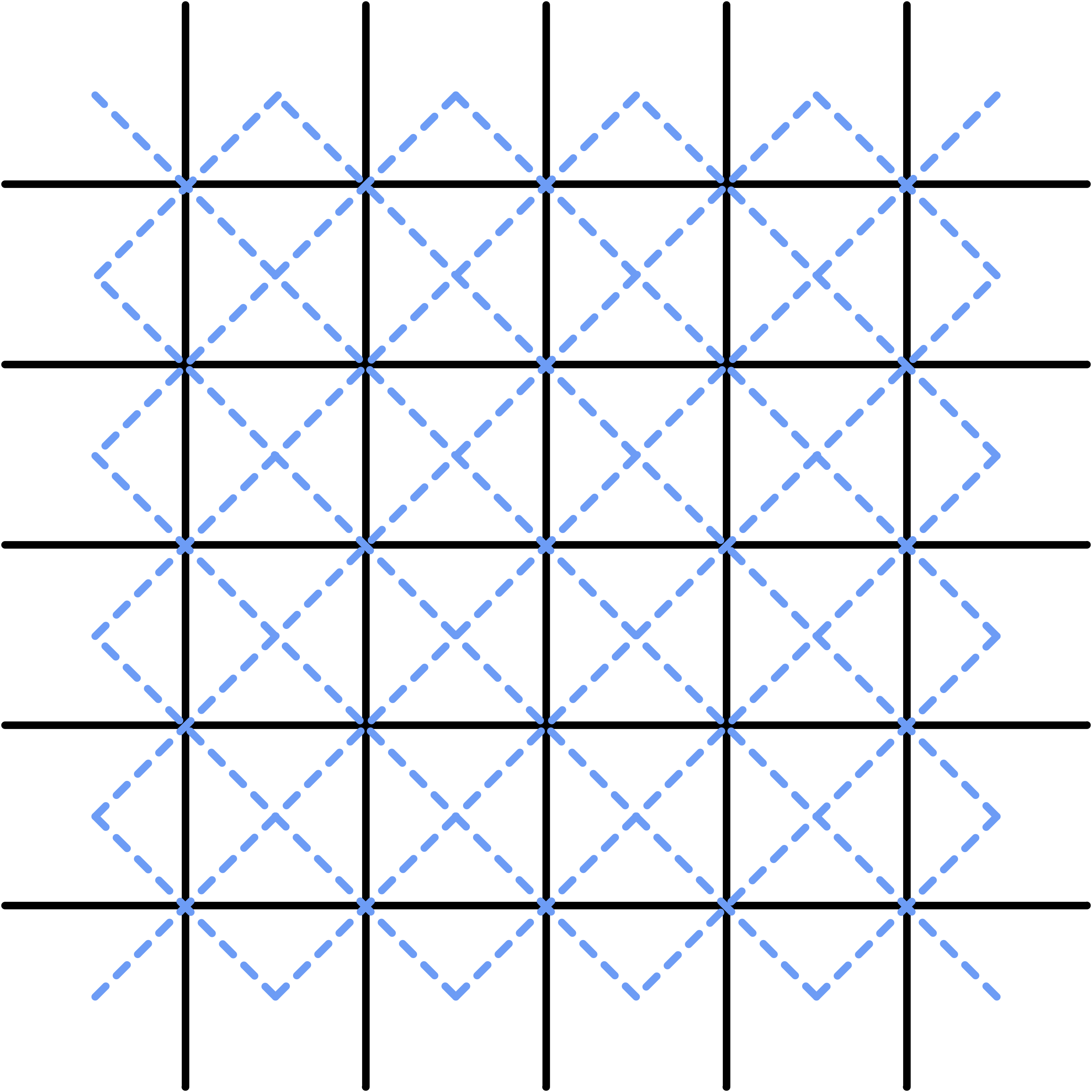}\caption{$L$}\label{fig:anti-ferro_lattice}
\end{subfigure}
\begin{subfigure}[b]{0.3\linewidth}
\centering\includegraphics[width=0.8\linewidth]{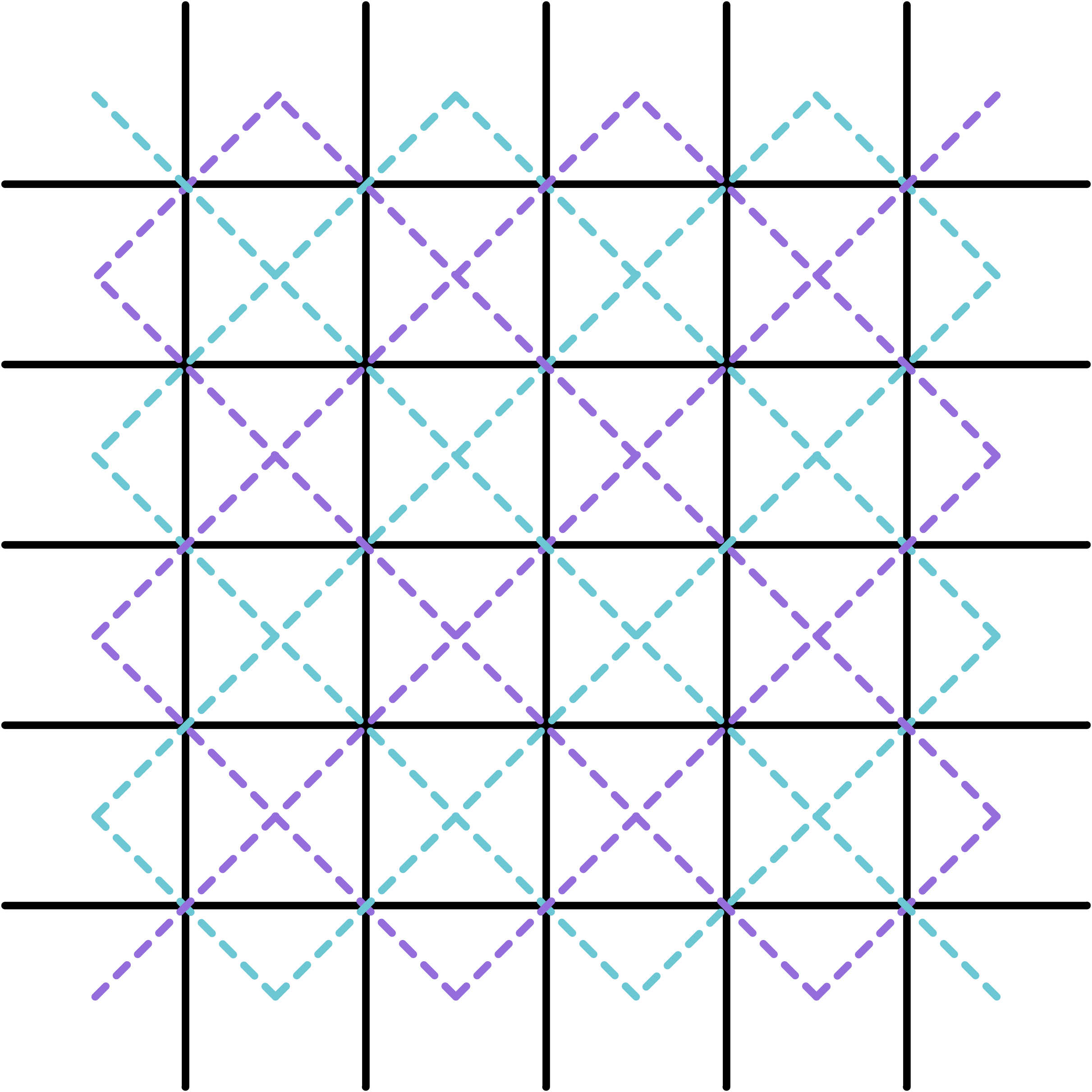}\caption{$L_0$ and $L_1$}
\label{fig:anti-ferro_lattice_sub}
\end{subfigure}
\begin{subfigure}[b]{0.3\linewidth}
\centering\includegraphics[width=0.8\linewidth]{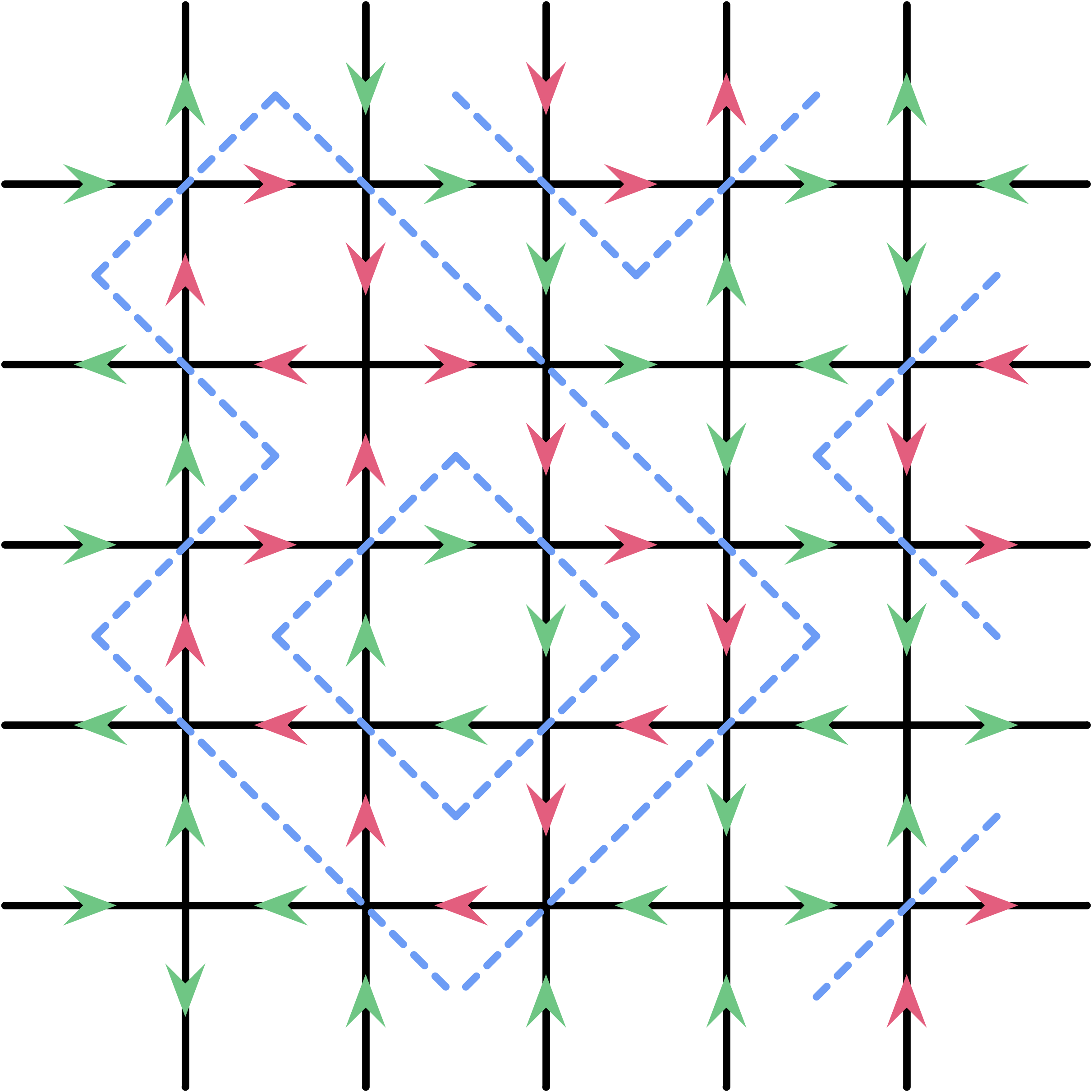}\caption{$\tau$ and $L_\tau$}
\label{fig:anti-ferro_eg_lattice}
\end{subfigure}
\caption{}\label{fig:anti-ferro_L}
\end{figure}

For a state $\tau \in \Omega$, we say $\tau$ has a \emph{horizontal (or vertical) fault line} if there is a self-avoiding path in $L_\tau$ connecting a $L_n$-vertex on the left (top, respectively) boundary of $L_n$ to a $L_n$-vertex on the right (bottom, respectively) boundary of $L_n$.
See \figref{fig:anti-ferro_tau_fault} for an example where a state has both a horizontal fault line and a vertical fault line.
Denote by $C_{\text{FL}}$ the set of states containing a horizontal fault line or a vertical fault line.
Since a fault line separates green edges from red edges, a vertical (horizontal) fault line precludes any horizontal (vertical, respectively) monochromatic bridge (the proof is basically the same as \lemref{lem:anti-ferro_no_cross}). This is to say, $C_\text{G}$, $C_\text{FL}$, and $C_\text{R}$ are pairwise disjoint.
Next we show the following lemma and its direct implication (\corref{cor:anti-ferro_2}).

\begin{lemma}\label{lem:anti-ferro_1}
If in a state $\tau$ there is no monochromatic cross, then there is a fault line.
\end{lemma}
\begin{proof}
If there is no monochromatic cross (i.e., a green cross or a red cross) in $\tau$, we can assume that 
\begin{equation} \label{cond:no_bridge}
\text{there is no green horizontal bridge and there is no red horizontal bridge.}\tag{$*$}
\end{equation}
\begin{itemize}
\item
Suppose $\tau$ has a green horizontal bridge. There is no red vertical bridge since it cannot ``cross'' the green horizontal bridge; there is no green vertical bridge since there is no green cross. Therefore, this case is symmetric to (\ref{cond:no_bridge}), switching horizontal for vertical.
\item
Suppose $\tau$ has a red horizontal bridge. This case is similar to the above case, and thus is also symmetric to (\ref{cond:no_bridge}).
\end{itemize}

Next we show there is a vertical fault line if there is no monochromatic horizontal bridge.
We introduce another graph $M_n$ that is the \emph{medial graph} of $\Lambda_n$ where every vertex of $M_n$ corresponds to an edge of $\Lambda_n$, i.e., two $M_n$-vertices are neighboring if the two corresponding $\Lambda_n$-edges are rotationally adjacent. Note that $M_n$ is part of another shifted square lattice. An example of $M_n$ and its relationship with $\mathbb{Z}^2$ is shown in \figref{fig:anti-ferro_medial}.
For any state $\tau \in \Omega$, there is a subset $M_\tau$ of $M_n$-edges associated with $\tau$. 
An $M_n$-edge $e$ is in $M_\tau$ if the two vertices that $e$ is incident to (as $\Lambda_n$-edges) have the same color (both green or both red). See \figref{fig:anti-ferro_tau_medial} for an example.

Observe that the correspondence between $\Lambda_n$-edges and $M_n$-vertices translates into the correspondence between simple monochromatic paths in $\Lambda_n$ to simple connected paths in $M_\tau$. In fact, the connected components in $M_\tau$, capturing the notion of monochromatic regions of $\Lambda_n$-edges, and connected components in $L_\tau$, capturing the notion of separation between regions of $\Lambda_n$-edges of different colors, are in a \emph{dual} relationship. 

This duality is depicted in \figref{fig:anti-ferro_tau_dual} and helps us find a fault line.
Let $V_M$ be the collection of $M_n$-vertices that can be reached from the left boundary of $M_n$ by a simple path in $M_\tau$. Since there is no monochromatic horizontal bridge, $V_M$ does not contain any $M_n$-vertex on the right boundary. As a consequence, there is a \emph{cutset} in $M_n$ separating $V_M$ from the right boundary. This cutset, composed of $M_n$-edges, corresponds to a vertical fault line.
For instance, in \figref{fig:anti-ferro_tau_dual} the blue solid $L_\tau$-path $\gamma$ is a fault line defined by the above argument.
\end{proof}
\begin{corollary}\label{cor:anti-ferro_2}
$\Omega = C_{\text{\textup{G}}} \cup C_\text{\textup{FL}}\cup C_{\text{\textup{R}}}$ is a partition of the state space.
\end{corollary}

\captionsetup[subfigure]{labelformat=parens}
\begin{figure}[h!]
\centering
\begin{subfigure}[b]{0.3\linewidth}
\centering\includegraphics[width=0.8\linewidth]{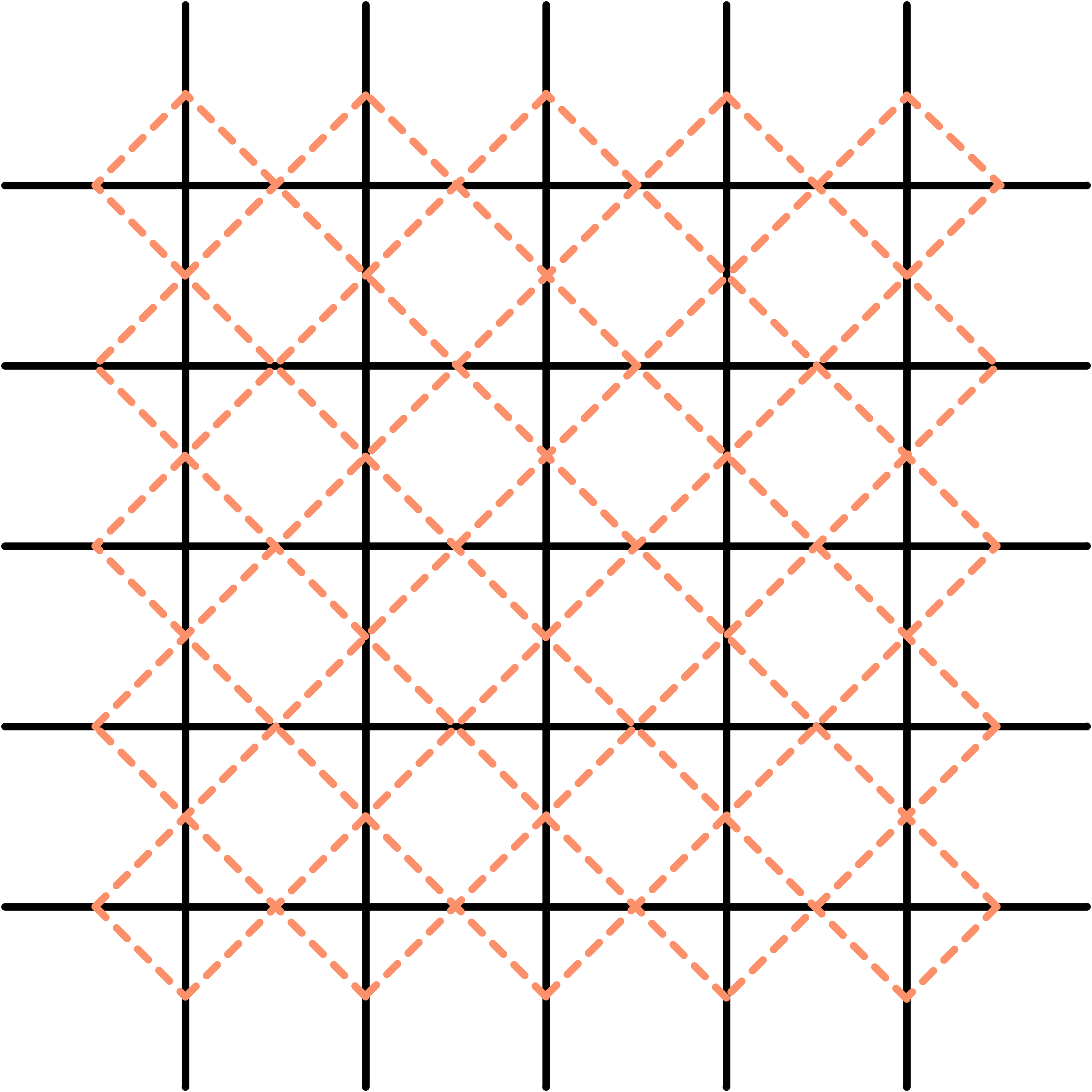}\caption{$M_n$}\label{fig:anti-ferro_medial}
\end{subfigure}
\begin{subfigure}[b]{0.3\linewidth}
\centering\includegraphics[width=0.8\linewidth]{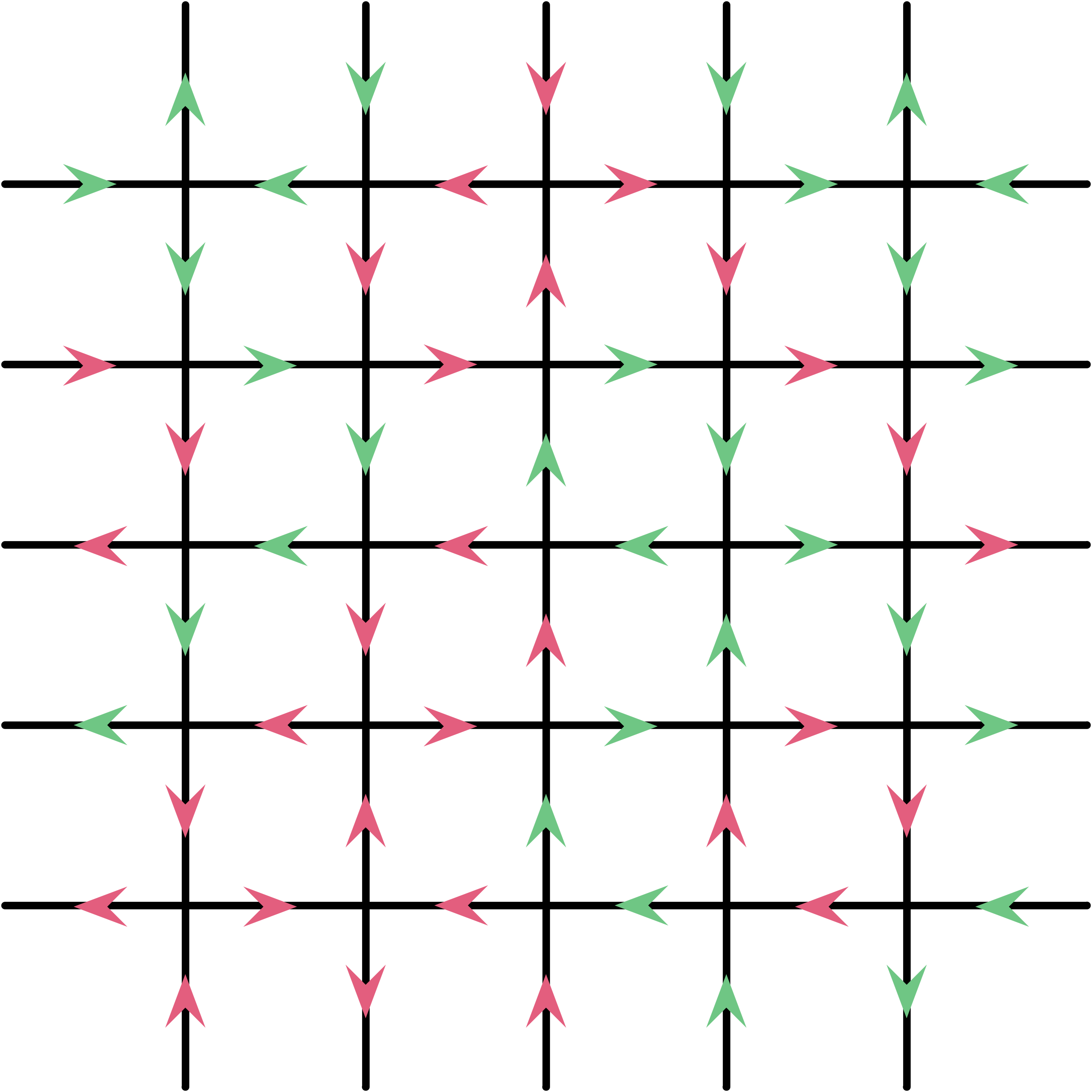}\caption{$\tau$}
\label{fig:anti-ferro_tau}
\end{subfigure}
\begin{subfigure}[b]{0.3\linewidth}
\centering\includegraphics[width=0.8\linewidth]{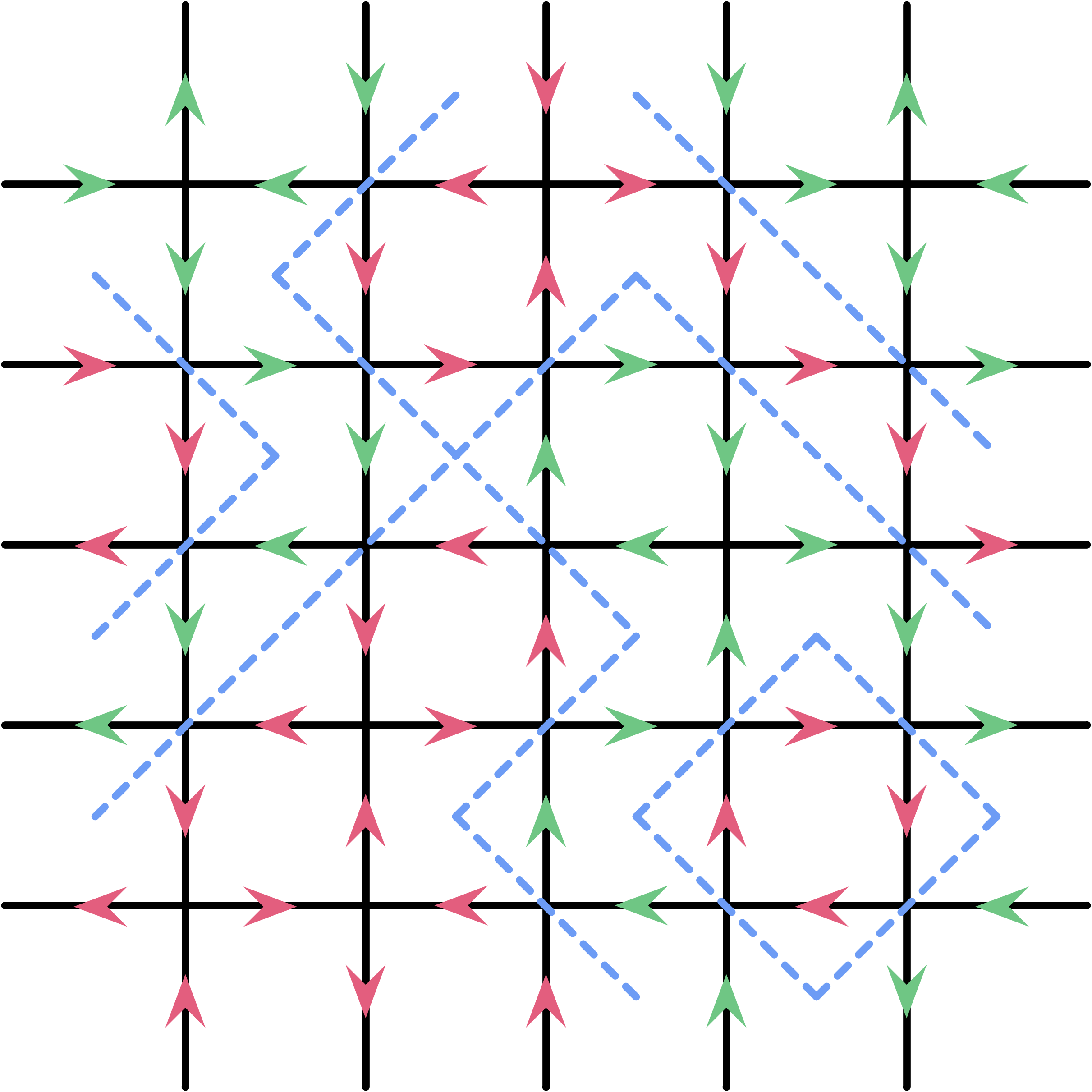}\caption{$L_\tau$}
\label{fig:anti-ferro_tau_fault}
\end{subfigure}
\begin{subfigure}[b]{0.3\linewidth}
\centering\includegraphics[width=0.8\linewidth]{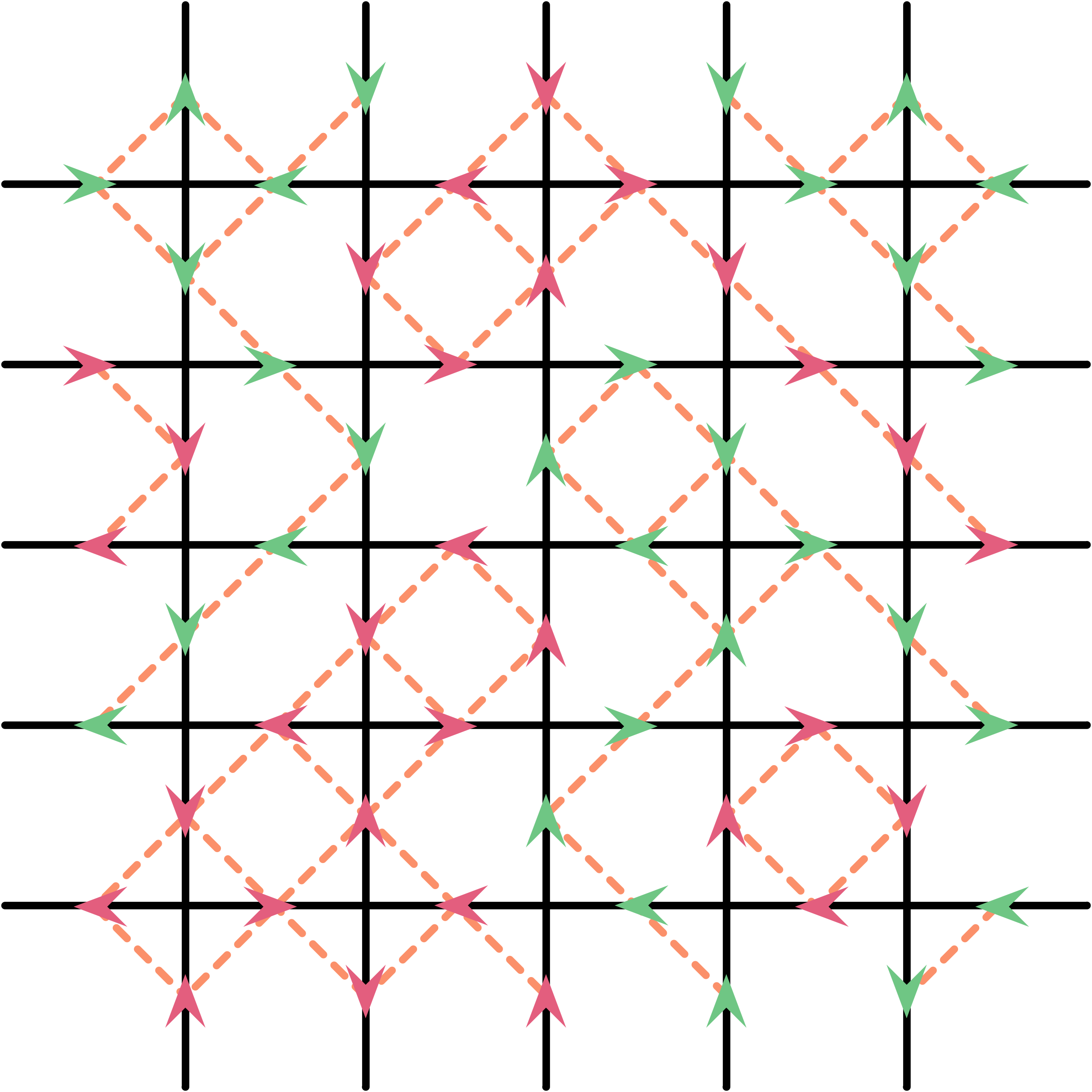}\caption{$M_\tau$}\label{fig:anti-ferro_tau_medial}
\end{subfigure}
\begin{subfigure}[b]{0.3\linewidth}
\centering\includegraphics[width=0.8\linewidth]{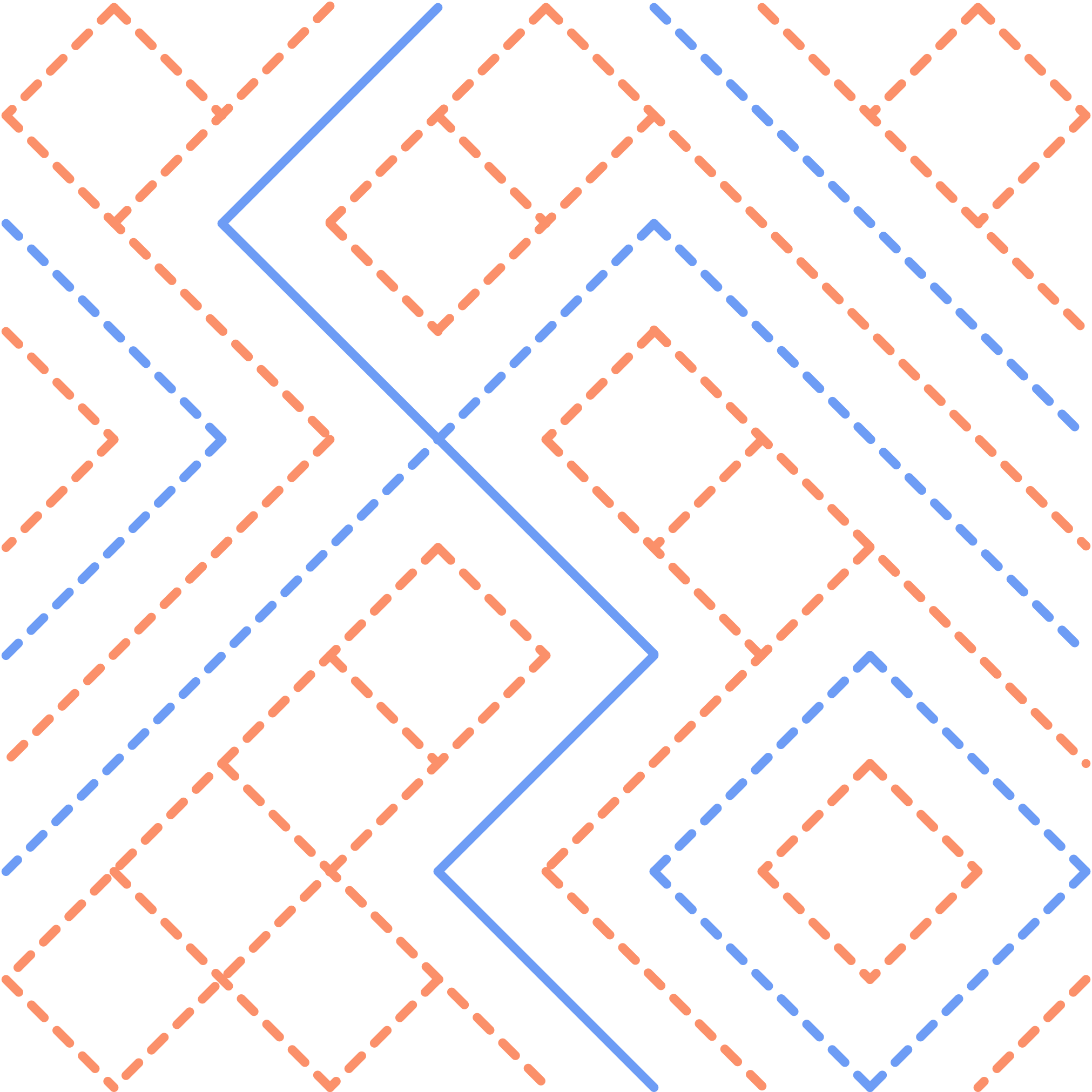}\caption{Duality}
\label{fig:anti-ferro_tau_dual}
\end{subfigure}
\begin{subfigure}[b]{0.3\linewidth}
\centering\includegraphics[width=0.8\linewidth]{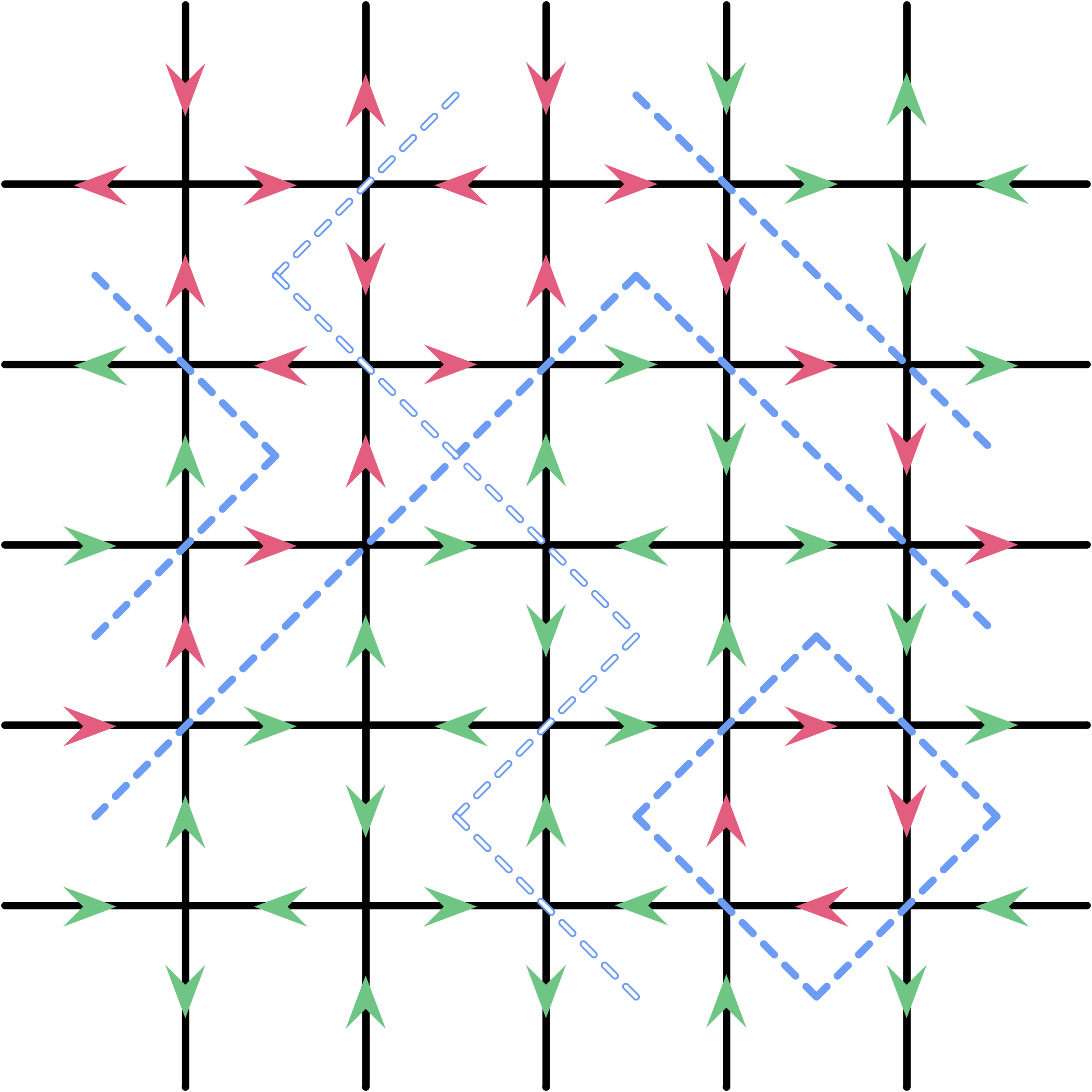}\caption{The injective map}
\label{fig:anti-ferro_tau_map}
\end{subfigure}
\caption{}\label{fig:anti-ferro_proof}
\end{figure}

Before moving on to prove \thmref{thm:anti-ferro_sub}, we introduce the notion of \emph{almost fault lines}.
A \emph{horizontal (or vertical) almost fault line} is a self-avoiding $L_n$-path connecting a $L_n$-vertex on the left boundary of $L_n$ to a $L_n$-vertex on the right boundary of $L_n$ where all edges except for one are in $L_\tau$. Denote by $C_\text{AFL}$ the set of states containing an almost fault line.
Let $\partial C_\text{G}$ be the set of states outside $C_\text{G}$ which are one-flip away from $C_\text{G}$ in the state space of $\mathcal{M}_G$.

\begin{lemma}\label{lem:anti-ferro_3}
$\partial C_\text{\textup{G}} \subset C_\text{\textup{FL}} \cup C_\text{\textup{AFL}}$.
\end{lemma}
\begin{proof}
If a state $\tau_\partial \in \partial C_\text{G}$ is not in $C_\text{FL}$ (does not have a fault line), then by \corref{cor:anti-ferro_2} $\tau_\partial \in C_\text{R}$ since by definition it is outside of $C_\text{G}$. Because $\tau_\partial$ is one move away from $C_\text{G}$, there exists a state $\tau_\text{G} \in C_\text{G}$ such that flipping four monochromatic edges along a unit square $s$ on $\mathbb{Z}^2$ yields $\tau_\partial$.
We know that in $\tau_\text{G}$ there exists a green cross and no red cross; in $\tau_\partial$ there exists a red cross and no green cross.
Then it must be true that the four edges along $s$ are all green in $\tau_\text{G}$ and all red in $\tau_\partial$.
See \figref{fig:anti-ferro_move} for a pictorial illustration.
Moreover, any green cross in $\tau_\text{G}$ must contain edges along $s$ and so is any red cross in $\tau_\partial$; otherwise, a green horizontal (vertical) bridge in $\tau_\text{G}$ must go across a red vertical (or horizontal, respectively) bridge in $\tau_\partial$ at some other place on $\Lambda_n$, which is impossible.

Therefore, there exists a simple green path $\Gamma_\text{G}$ from some vertices on $s$ to the top boundary of $\Lambda_n$ and a simple red path $\Gamma_\text{R}$ from some vertices on $s$ to the top boundary of $\Lambda_n$. In the following, we prove that the above conditions suffice to show that there exists an $L_\tau$-path from $s$ to the top boundary of $L_n$. Similar conclusions can be made for the existence of $L_\tau$-paths from $s$ to the bottom boundary of $L_n$. By adding at most one $L_n$-edge, we can concatenate these paths to obtain a vertical almost fault line.

$\Gamma_\text{G}$ and $\Gamma_\text{R}$ cannot cross each other (\figref{fig:anti-ferro_no_cross}). Without loss of generality, suppose $\Gamma_\text{G}$ is to the left of $\Gamma_\text{R}$.
Then we use the medial lattice view $M_{\tau_\partial}$ as in the proof of \lemref{lem:anti-ferro_1}. $\Gamma_\text{G}$ corresponds to a connected component $m_\text{G}$ in $M_{\tau_\partial}$.
Denote by $V_\text{G}$ the set of $M_n$-vertices which can be reached by $m_\text{G}$ in $M_{\tau_\partial}$, and let $V'_\text{G}$ be $V_\text{G}$ together with the set of $M_n$-vertices which are separated from the right boundary of $M_n$ by $V_\text{G}$.
Then there is a \emph{cutset} in $M_n$ separating $V'_\text{G}$ from the right boundary. This cutset, composed of $M_n$-edges, corresponds to a dual $L_\tau$-path from $s$ to the top boundary of $L_n$.
\end{proof}

\captionsetup[subfigure]{labelformat=parens}
\begin{figure}[h!]
\centering
\begin{subfigure}[b]{0.4\linewidth}
\centering\includegraphics[width=0.5\linewidth]{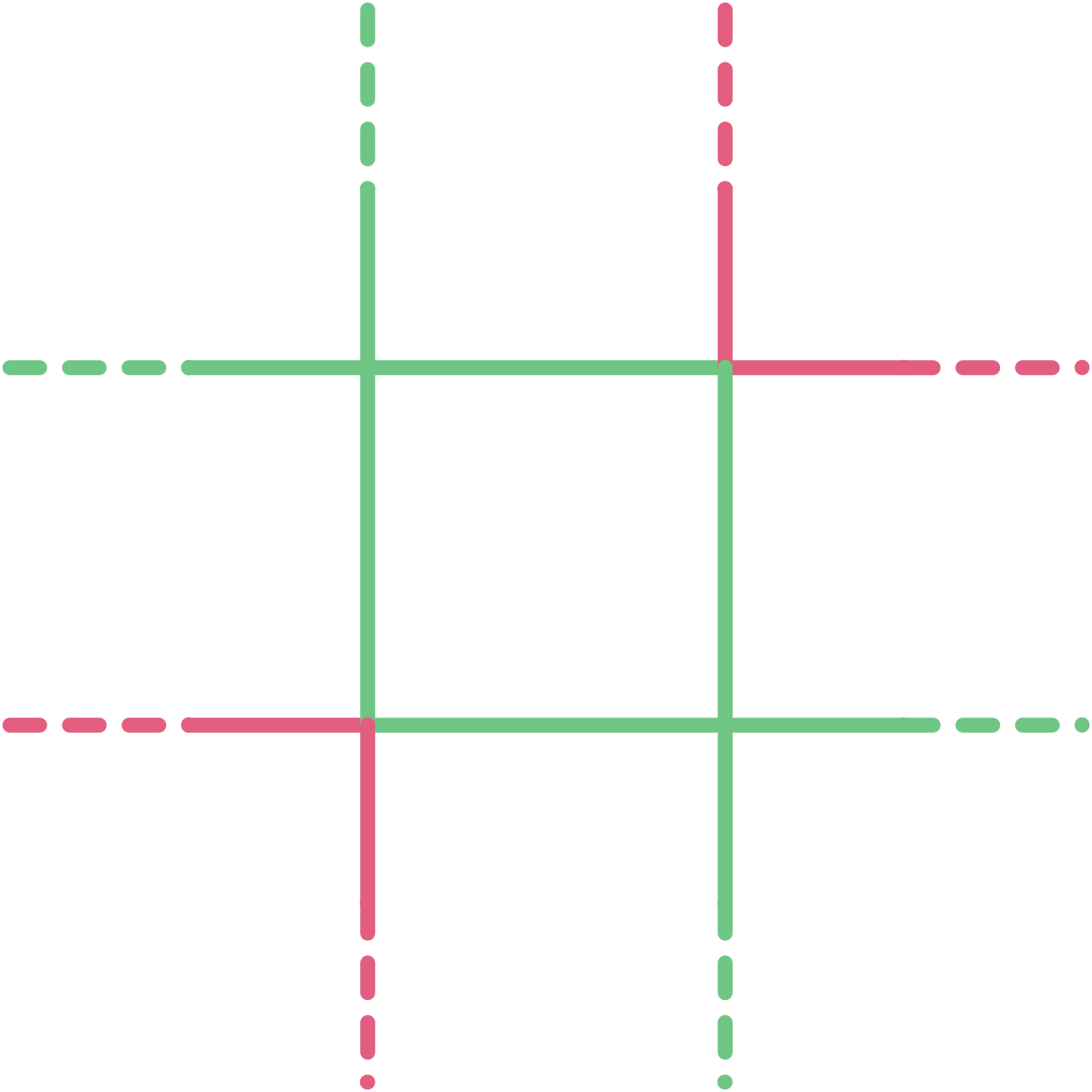}\caption{$\tau_\text{G}$}\label{fig:anti-ferro_move_1}
\end{subfigure}
\begin{subfigure}[b]{0.4\linewidth}
\centering\includegraphics[width=0.5\linewidth]{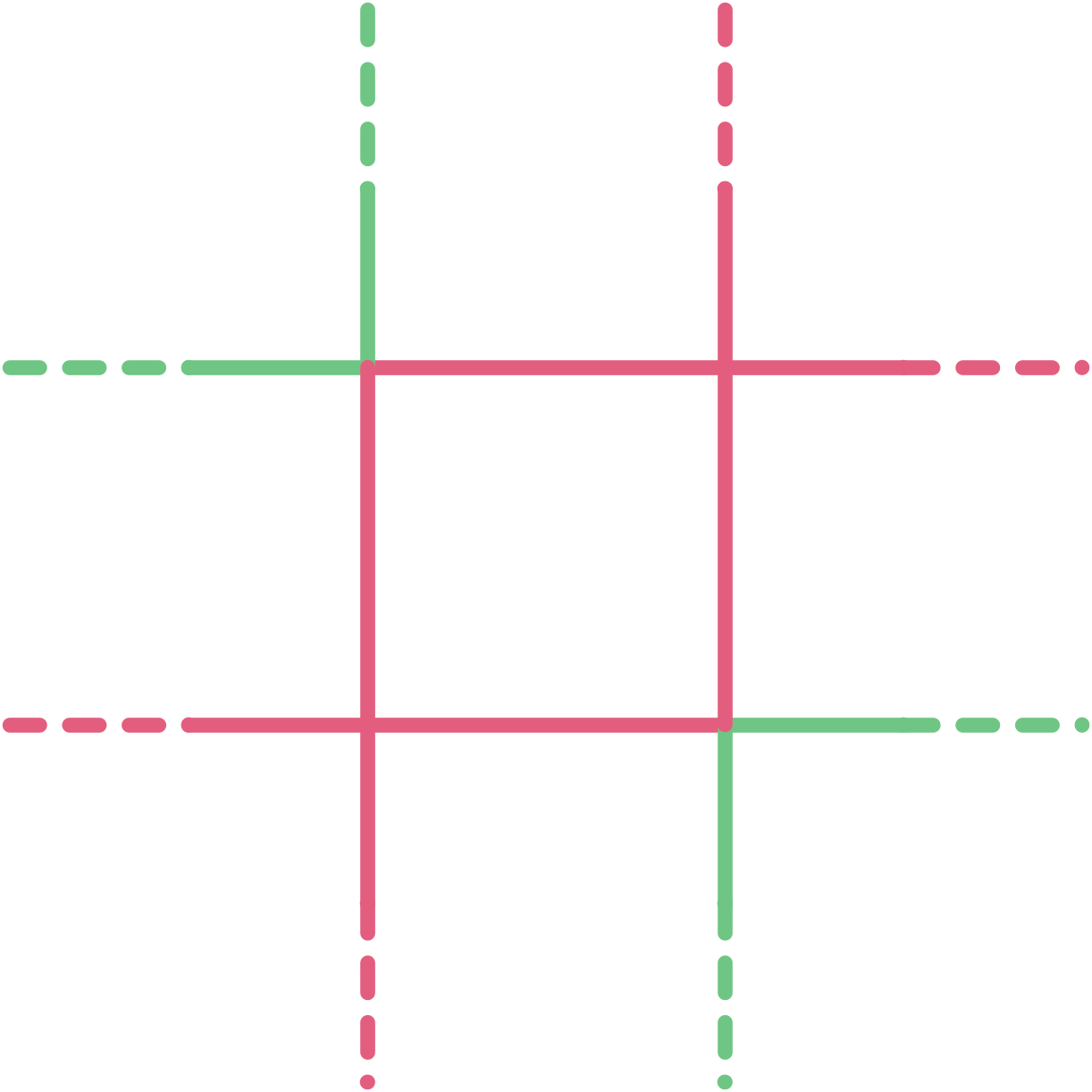}\caption{$\tau_\partial$}
\label{fig:anti-ferro_move_2}
\end{subfigure}
\caption{A step in $\mathcal{M}_G$.}\label{fig:anti-ferro_move}
\end{figure}

Now we are ready to show \thmref{thm:anti-ferro_sub}.
Let $\Omega_\text{LEFT} = C_\text{G}$, $\Omega_\text{MIDDLE} = C_\text{FL} \cup (C_\text{AFL} \cap C_\text{R})$, and $\Omega_\text{RIGHT} = C_\text{R} \setminus C_\text{AFL}$.
\thmref{thm:anti-ferro_sub} is a consequence of the following lemma which uses a \emph{Peierls argument} to show that $\pi(C_\text{FL} \cup C_\text{AFL})$ is exponentially small.

\begin{lemma}
$\pi(C_\text{\textup{FL}} \cup C_\text{\textup{AFL}}) \le O(n)\left(\frac{2.639\max(a,b)}{c}\right)^n$.
\end{lemma}
\begin{proof}
For a self-avoiding path $\gamma$ in $L_n$ connecting a vertex on the top boundary to a vertex on the bottom boundary, denote by $F_\gamma$ the set of states in $\Omega$ that contain $\gamma$ as vertical fault line or almost fault line.
Reversing directions of all the edges to the left side of $\gamma$ defines an injective mapping from $F_\gamma$ to $\Omega \setminus F_\gamma$ that magnifies probability by a factor of at least $\frac{\min(a,b)}{c} \cdot \left(\frac{c}{\max(a,b)}\right)^{|\gamma|-1}$.
This is because: if $\gamma$ is a fault line in a state, every $\Lambda_n$-vertex sitting on $\gamma$ would have four incident edges in the same color after the map, which increase its weight to $c$ (from $a$ or $b$); if $\gamma$ is an almost fault line in a state, the above is true except that for one $\Lambda_n$-vertex, its weight decrease from $c$ to $a$ or $b$ after the map, as orientations on half of the four monochromatic edges are reversed.
This indicates that $\pi(F_\gamma) \le \frac{c}{\min(a,b)} \cdot \left(\frac{\max(a,b)}{c}\right)^{|\gamma|-1}$. See \figref{fig:anti-ferro_tau_map} for an example.
The same goes for horizontal (almost) fault lines.

Since every (almost) fault line is a self-avoiding walk on $L$, the number of fault lines of length $l$ is upper bounded by $2n$ times the number of self-avoiding walks of that length starting at a vertex on the left or bottom boundary. The latter can be bounded by an well-studied estimate $\mu^l$ on the number of self-avoiding walks of length $l$ on $\mathbb{Z}^2$, where $\mu$ is called the \emph{connective constant}. The best proved bound is $\mu \approx 2.638158\cdots$~\cite{Guttmann2001}.
Summing this over fault lines of length from $n$ to $n^2$ completes the proof.
\end{proof}

We have proved the torpid mixing of Glauber dynamics for the six-vertex model on the lattice region $\Lambda_n$ with free boundary conditions. Next we state the idea to extend the proof for the case when the Markov chain is $\mathcal{M}_D$ and
the case when the boundary of $\widetilde{\Lambda}_n$ is periodic.
% and side length is even.
\thmref{thm:anti-ferro} is a combination of \thmref{thm:anti-ferro_sub} and the extensions.

\begin{figure}[h!]
\centering
\includegraphics[width=0.24\linewidth]{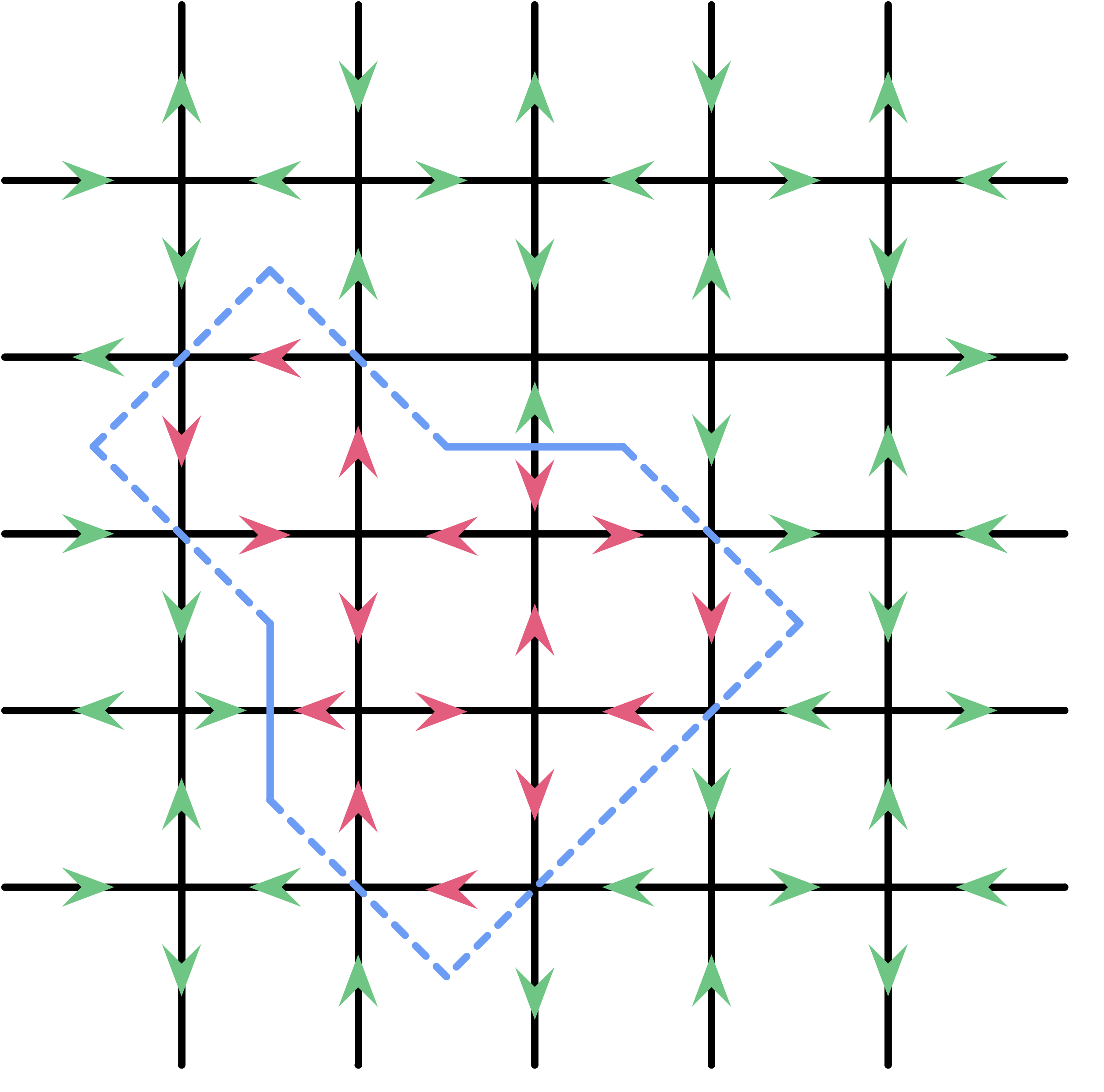}
\caption{A state in $\mathcal{M}_D$.}\label{fig:anti-ferro_defects}
\end{figure}

To extend \thmref{thm:anti-ferro_sub} to hold for the directed-loop algorithm $\mathcal{M_D}$ whose state space is $\Omega \cup \Omega'$, we need to pay extra attention for the states in $\Omega'$, the near-perfect Eulerian orientations. For a state $\tau' \in \Omega'$, there are two ``defects'' on the edges (\figref{fig:anti-ferro_defects}). Apart from the diagonal $L_n$-edges, there are two $\left(\mathbb{Z}^2 + \left(\frac{1}{2}, \frac{1}{2}\right)\right)$-edges separating green (half-)edges from red (half-)edges. The adaption we make is to put such $\left(\mathbb{Z}^2 + \left(\frac{1}{2}, \frac{1}{2}\right)\right)$-edges also into the set $L_{\tau'}$. Notice that a connected component in $L_{\tau'}$ could possibly lie on $L_0$-edges as well as $L_1$-edges.
Everything we prove is still correct if we allow fault lines to have $\left(\mathbb{Z}^2 + \left(\frac{1}{2}, \frac{1}{2}\right)\right)$-edges. Due to the possible positions of such two defects, the weight of $C_\text{FL} \cup C_\text{AFL}$ only increases by a polynomial factor in $n$, hence not affecting the fact of torpid mixing.

%To extend \thmref{thm:anti-ferro_sub} to hold for $\widetilde{\Lambda}_n$ with periodic boundary condition (i.e., a 2-dimensional torus), our proof is similar to the treatment in \cite{Randall:2006:SMG:1109557.1109653} for the hardcore gas model on the square lattice regions with periodic boundary conditions.
To extend \thmref{thm:anti-ferro_sub} to hold for $\widetilde{\Lambda}_n$ with periodic boundary condition (i.e., a 2-dimensional torus), we make the following modification.
%, similar to the treatment in \cite{Randall:2006:SMG:1109557.1109653} for the hardcore gas model on the square lattice.
When $n$ is even, there still are two states with maximum weights $c^{n^2}$ (similar to $\tau_\text{G}$ and $\tau_\text{R}$ in \figref{fig:anti-ferro}). Again, for any state $\tau$, $\widetilde{\Lambda}_n$-edges can be classified as green or red, and its associated $L_\tau$ separates $\widetilde{\Lambda}_n$-edges of different colors.
For the 2-dimensional torus $T^2$, the homology group $H_1(T^2) \cong \mathbb{Z} \times \mathbb{Z}$.
We say a state $\tau$ has a \emph{green (or red) cross} if there are two %\emph{non-self-intersecting}
\emph{non-contractable} cycles of green (or red, respectively) edges %in different \emph{homology classes};
of \emph{homology classes} $(a_1, b_1)$ and $(a_2, b_2)$ with $\operatorname{det}\left[\begin{smallmatrix} a_1 a_2 \\ b_1 b_2 \end{smallmatrix}\right] \neq 0$;
$\tau$ has \emph{a pair of} fault lines if there is a pair of non-contractable cycles of $L_\tau$-edges. (By parity, if there is one $L_\tau$-cycle there must be two.)
Then the proofs in this section can be naturally adapted for the torus case, and the torpid mixing result follows.

\section*{Acknowledgement}
The author thanks Professor Jin-Yi Cai for the valuable suggestions on the preliminary version of this paper.

\bibliography{reference}{}

\newcommand{\etalchar}[1]{$^{#1}$}
\begin{thebibliography}{BCK{\etalchar{+}}99}

\bibitem[AR05]{Allison2005}
David Allison and Nicolai Reshetikhin.
\newblock Numerical study of the 6-vertex model with domain wall boundary
  conditions.
\newblock {\em Annales de l'Institut Fourier}, 55(6):1847--1869, 2005.

\bibitem[Bax82]{Baxter:book}
R.~J. Baxter.
\newblock {\em Exactly Solved Models in Statistical Mechanics}.
\newblock Academic Press, 1982.

\bibitem[BCK{\etalchar{+}}99]{Borgs:1999:TMM:795665.796518}
Christian Borgs, Jennifer~T. Chayes, Jeong~Han Kim, Alan Frieze, Prasad Tetali,
  Eric Vigoda, and Van~Ha Vu.
\newblock Torpid mixing of some {Monte} {Carlo} {Markov} chain algorithms in
  statistical physics.
\newblock In {\em Proceedings of the 40th Annual Symposium on Foundations of
  Computer Science (FOCS)}, pages 218--229, 1999.

\bibitem[BGRT13]{DBLP:conf/approx/BlancaGRT13}
Antonio Blanca, David Galvin, Dana Randall, and Prasad Tetali.
\newblock Phase coexistence and slow mixing for the hard-core model on
  {$\mathbb{Z}^2$}.
\newblock In {\em Approximation, Randomization, and Combinatorial Optimization.
  Algorithms and Techniques (APPROX/RANDOM)}, pages 379--394, 2013.

\bibitem[BKW73]{doi:10.1063/1.1666271}
H.~J. Brascamp, H.~Kunz, and F.~Y. Wu.
\newblock Some rigorous results for the vertex model in statistical mechanics.
\newblock {\em Journal of Mathematical Physics}, 14(12):1927--1932, 1973.

\bibitem[BN98]{PhysRevE.57.1155}
G.~T. Barkema and M.~E.~J. Newman.
\newblock {Monte Carlo} simulation of ice models.
\newblock {\em Phys. Rev. E}, 57:1155--1166, Jan 1998.

\bibitem[CGMS96]{Cesi1996}
F.~Cesi, G.~Guadagni, F.~Martinelli, and R.~H. Schonmann.
\newblock On the two-dimensional stochastic {Ising} model in the phase
  coexistence region near the critical point.
\newblock {\em Journal of Statistical Physics}, 85(1):55--102, Oct 1996.

\bibitem[CLL17]{DBLP:journals/corr/abs-1712-05880}
Jin{-}Yi Cai, Tianyu Liu, and Pinyan Lu.
\newblock Approximability of the six-vertex model.
\newblock {\em CoRR}, abs/1712.05880, 2017.

\bibitem[{Elo}99]{1999JSP....96.1091E}
K.~{Eloranta}.
\newblock {Diamond Ice}.
\newblock {\em Journal of Statistical Physics}, 96:1091--1109, September 1999.

\bibitem[FW70]{PhysRevB.2.723}
Chungpeng Fan and F.~Y. Wu.
\newblock General lattice model of phase transitions.
\newblock {\em Phys. Rev. B}, 2:723--733, Aug 1970.

\bibitem[GC01]{Guttmann2001}
A.J. Guttmann and A.R. Conway.
\newblock Square lattice self-avoiding walks and polygons.
\newblock {\em Annals of Combinatorics}, 5(3):319--345, Dec 2001.

\bibitem[GMP04]{RSA:RSA20002}
Leslie~Ann Goldberg, Russell Martin, and Mike Paterson.
\newblock Random sampling of 3-colorings in $\mathbb{Z}^2$.
\newblock {\em Random Structures \& Algorithms}, 24(3):279--302, 2004.

\bibitem[Lie67a]{PhysRevLett.18.1046}
Elliott~H. Lieb.
\newblock Exact solution of the {$F$} model of an antiferroelectric.
\newblock {\em Phys. Rev. Lett.}, 18:1046--1048, Jun 1967.

\bibitem[Lie67b]{PhysRevLett.19.108}
Elliott~H. Lieb.
\newblock Exact solution of the two-dimensional {Slater} {KDP} model of a
  ferroelectric.
\newblock {\em Phys. Rev. Lett.}, 19:108--110, Jul 1967.

\bibitem[Lie67c]{PhysRev.162.162}
Elliott~H. Lieb.
\newblock Residual entropy of square ice.
\newblock {\em Phys. Rev.}, 162:162--172, Oct 1967.

\bibitem[LKV17]{1742-5468-2017-5-053103}
I~Lyberg, V~Korepin, and J~Viti.
\newblock The density profile of the six vertex model with domain wall boundary
  conditions.
\newblock {\em Journal of Statistical Mechanics: Theory and Experiment},
  2017(5):053103, 2017.

\bibitem[LPW06]{LevinPeresWilmer2006}
David~A. Levin, Yuval Peres, and Elizabeth~L. Wilmer.
\newblock {\em {Markov} chains and mixing times}.
\newblock American Mathematical Society, 2006.

\bibitem[LRS01]{doi:10.1137/S0097539799360355}
Michael Luby, Dana Randall, and Alistair Sinclair.
\newblock {Markov} chain algorithms for planar lattice structures.
\newblock {\em SIAM Journal on Computing}, 31(1):167--192, 2001.

\bibitem[LS12]{Lubetzky2012}
Eyal Lubetzky and Allan Sly.
\newblock Critical {Ising} on the square lattice mixes in polynomial time.
\newblock {\em Communications in Mathematical Physics}, 313(3):815--836, Aug
  2012.

\bibitem[MO94a]{Martinelli1994I}
F.~Martinelli and E.~Olivieri.
\newblock Approach to equilibrium of {Glauber} dynamics in the one phase
  region. {I}. the attractive case.
\newblock {\em Communications in Mathematical Physics}, 161(3):447--486, Apr
  1994.

\bibitem[MO94b]{Martinelli1994II}
F.~Martinelli and E.~Olivieri.
\newblock Approach to equilibrium of {Glauber} dynamics in the one phase
  region. {II}. the general case.
\newblock {\em Communications in Mathematical Physics}, 161(3):487--514, Apr
  1994.

\bibitem[MW96]{Mihail1996}
M.~Mihail and P.~Winkler.
\newblock On the number of {Eulerian} orientations of a graph.
\newblock {\em Algorithmica}, 16(4):402--414, Oct 1996.

\bibitem[Pau35]{doi:10.1021/ja01315a102}
Linus Pauling.
\newblock The structure and entropy of ice and of other crystals with some
  randomness of atomic arrangement.
\newblock {\em Journal of the American Chemical Society}, 57(12):2680--2684,
  1935.

\bibitem[Pei36]{peierls_1936}
R.~Peierls.
\newblock Statistical theory of adsorption with interaction between the
  adsorbed atoms.
\newblock {\em Mathematical Proceedings of the Cambridge Philosophical
  Society}, 32(3):471--476, 1936.

\bibitem[Ran06]{Randall:2006:SMG:1109557.1109653}
Dana Randall.
\newblock Slow mixing of {Glauber} dynamics via topological obstructions.
\newblock In {\em Proceedings of the Seventeenth Annual ACM-SIAM Symposium on
  Discrete Algorithm (SODA)}, pages 870--879, 2006.

\bibitem[RS72]{doi:10.1063/1.1678874}
Aneesur Rahman and Frank~H. Stillinger.
\newblock Proton distribution in ice and the {Kirkwood} correlation factor.
\newblock {\em The Journal of Chemical Physics}, 57(9):4009--4017, 1972.

\bibitem[RT00]{doi:10.1063/1.533199}
Dana Randall and Prasad Tetali.
\newblock Analyzing {Glauber} dynamics by comparison of {Markov} chains.
\newblock {\em Journal of Mathematical Physics}, 41(3):1598--1615, 2000.

\bibitem[SJ89]{SINCLAIR198993}
Alistair Sinclair and Mark Jerrum.
\newblock Approximate counting, uniform generation and rapidly mixing {Markov}
  chains.
\newblock {\em Information and Computation}, 82(1):93--133, 1989.

\bibitem[Sut67]{PhysRevLett.19.103}
Bill Sutherland.
\newblock Exact solution of a two-dimensional model for hydrogen-bonded
  crystals.
\newblock {\em Phys. Rev. Lett.}, 19:103--104, Jul 1967.

\bibitem[SZ04]{PhysRevE.70.016118}
Olav~F. Sylju\aa{}sen and M.~B. Zvonarev.
\newblock Directed-loop {Monte Carlo} simulations of vertex models.
\newblock {\em Phys. Rev. E}, 70:016118, Jul 2004.

\bibitem[YN79]{YANAGAWA1979329}
A.~Yanagawa and J.F. Nagle.
\newblock Calculations of correlation functions for two-dimensional square ice.
\newblock {\em Chemical Physics}, 43(3):329--339, 1979.

\end{thebibliography}
\bibliographystyle{alpha}

\end{document}